\documentclass[aps,prd,groupedaddress,nofootinbib,preprintnumbers,notitlepage]{revtex4-1}
\pdfoutput = 1
\usepackage{graphicx}
\usepackage{latexsym}
\usepackage{amsfonts}
\usepackage{amssymb}
\usepackage{amsmath}
\usepackage{slashed}
\usepackage{tabularx,url,color}
\usepackage{hyperref}
\hypersetup{colorlinks}

\definecolor{rot}{cmyk}{0,1,1,0.55}
\definecolor{blau}{cmyk}{1,1,0,0.6}
\definecolor{gruen}{cmyk}{0.8,0,0.5,0.2}

\hypersetup{colorlinks,bookmarksopen,bookmarksnumbered,
linkcolor=blau,urlcolor=rot,citecolor=gruen}

\usepackage{multirow}

\usepackage{booktabs}

\makeatletter
\newcommand\footnoteref[1]{\protected@xdef\@thefnmark{\ref{#1}}\@footnotemark}
\makeatother

\begin{document}

\title{Boosted Higgs Shapes}

\preprint{IPPP/14/35}
\preprint{DCPT/14/70}
\preprint{CERN-PH-TH/2014-083}
\preprint{DESY/14-069}
\preprint{KCL-PH-TH/2014-22}

\author{Matthias Schlaffer$^1$, Michael Spannowsky$^2$, Michihisa Takeuchi$^3$, Andreas Weiler$^{1,4}$, and Chris Wymant$^{2,5,}$\footnote{Present address:
   Department of Infectious Disease Epidemiology,
   Imperial College London,
   St Mary's Campus,
   Norfolk Place
   London, W2 1PG, UK.}}

\affiliation{$^1$DESY, Notkestrasse 85, D-22607 Hamburg, Germany}
\affiliation{$^2$Institute for Particle Physics Phenomenology, Department of Physics, Durham University, Durham DH1 3LE, UK}
\affiliation{$^3$Theoretical Physics and Cosmology Group, Department of Physics, King's College London, London~WC2R 2LS, UK}
\affiliation{$^4$Theory Division, Physics Department, CERN, CH-1211 Geneva 23, Switzerland}
\affiliation{$^5$Laboratoire d'Annecy-le-Vieux de Physique Th\'{e}orique, 9 Chemin de Bellevue, 74940 Annecy-le-Vieux, France}
\date{\today}

\begin{abstract}

The inclusive Higgs production rate through gluon fusion has been measured to be in agreement with the Standard Model (SM). We show that even if the inclusive 
Higgs production rate is very SM-like, a precise determination of the boosted Higgs transverse momentum shape offers the opportunity 
to see effects of natural new physics. These measurements are generically motivated by effective field theory arguments and specifically in
extensions of the SM with a natural weak scale, like
 composite Higgs models and natural supersymmetry. We show in detail how a measurement at high transverse momentum 
of $H\to2\ell+\slashed{\mathbf{p}}_T$ via $H\to \tau\tau$ and $H\to WW^*$ 
could be performed and demonstrate that it offers a compelling alternative to the $t\bar t H$ channel. We 
discuss the sensitivity to new physics in the most challenging scenario of an exactly SM-like inclusive Higgs cross-section.

\end{abstract}

\maketitle

\section{Introduction}
\label{sec:intro}

After the observation of a Standard Model-like Higgs
boson~\cite{Aad:2012tfa,Chatrchyan:2012ufa} the main physics program has shifted to measuring its
precise properties. 
The exotic possibility of the new particle having spin one or two is already disfavored~\cite{Aad:2013xqa,Bolognesi:2012mm} by the
analysis of its decays into $\gamma\gamma$~\cite{ATLAS-CONF-2013-029,CMS-PAS-HIG-13-016, Landau:1948kw,Yang:1950rg},
$WW^*$~\cite{ATLAS:2013vla,CMS-PAS-HIG-13-003,Ellis:2012jv}, and
$ZZ^*$~\cite{ATLAS-CONF-2013-013,Chatrchyan:2013mxa,Bolognesi:2012mm,Choi:2002jk,Buszello:2002uu,Bredenstein:2006rh,Cao:2009ah,Gao:2010qx,Englert:2010ud, Ellis:2012wg,Ellis:2012mj}; so
is the possibility of it being a pure pseudo-scalar~\cite{Freitas:2012kw,Coleppa:2012eh}, though it could still be an
admixture of scalar and pseudo-scalar~\cite{Plehn:2001nj,Klamke:2007cu,Godbole:2007cn,Berge:2011ij,Englert:2012ct, Djouadi:2013qya,Harnik:2013aja,Ellis:2013yxa}.  
The couplings to gauge bosons and fermions have also been measured for various decay modes, and they are so far consistent with the SM predictions
within uncertainties~\cite{ATLAS-CONF-2014-009,Chatrchyan:2014vua}.\par

With no apparent deviation from the SM so far, it is important to closely examine the channels where
one has a fighting chance to encounter new physics. One such promising process is Higgs production via
gluon fusion: In order to avoid unnatural fine-tuning while still obtaining a light Higgs mass, 
loops of new particles need to soften to the Higgs mass squared UV sensitivity of the top loop.
If these particles are charged under the \(SU(3)\) color gauge
group (which they are in almost all known cases), gluons will couple to the loop.
With two gluons coupled to the new physics loop and one Higgs set to its
vacuum expectation value, one gets a contribution to gluon fusion, 
the dominant Higgs production mechanism, see e.g.~\cite{Low:2009di,Espinosa:2012in}.

At the same time, top partners can lead to a modified top-Yukawa coupling. A change in the top-Yukawa 
affects the Higgs production cross-section and can even compensate for
new particles in the loop such that a SM-like inclusive cross-section is
obtained even though new physics is present. The reason for this is that already for the top
quark the effective gluon Higgs interaction \cite{Shifman:1979eb, Ellis:1975ap} obtained from the
low energy theorem is a very good description \cite{Kniehl:1995tn,Grazzini:2013mca} which works even better for
heavier particles. Therefore the inclusive amplitude can be expressed as the sum of two identical
Feynman diagrams with the effective interaction (one from the top loop and one from the non-SM loop)
which differ only by a coefficient, \(c_t\) and \(\kappa_g\), respectively. The cross-section is
therefore only sensitive to the absolute square of the sum of these coefficients. The effects of
this in composite-Higgs models were calculated
in~\cite{Falkowski:2007hz,Low:2010mr,Azatov:2011qy,Montull:2013mla,Delaunay:2013iia} where it is
shown that the contributions to the inclusive cross-section indeed cancel in minimal models.\par

The main idea now is to study boosted Higgs shapes above a certain $p_T$ scale.
This scale should be high enough to resolve the top loop beyond the effective description, but 
low enough to keep the effective description of the loop of the new particle valid; see~\cite{Grojean:2013nya}
for a discussion in a concrete model. The simple relation 
\(\sigma\propto\left|c_t+\kappa_g\right|^2\) for the inclusive cross-sections 
does not apply but is modified, allowing
the two coefficients to be extracted separately, when combined with the inclusive measurement. Early
studies looking for New Physics in the Higgs \(p_T\) distribution in the gluon-fusion production mode include \cite{Langenegger:2006wu,
  Arnesen:2008fb, Bagnaschi:2011tu} and recent
preliminary studies looking at highly boosted Higgs shapes
include~\cite{Grojean:2013nya,Azatov:2013xha,Harlander:2013oja,Banfi:2013yoa}.  An alternative
approach to measure the coupling \(c_t\) in boosted \(p p \to HZ\) was presented
in~\cite{Englert:2013vua, Harlander:2013mla}. Although there are attempts to measure \(c_t\) directly by looking into
the difficult \(t\bar{t}H\) channel~\cite{Plehn:2009rk, ATLAS2014ttH, CMS:2013fda,CMS:2013sea,CMS:2013tfa,Artoisenet:2013vfa,Buckley:2013auc},
it is important to explore
boosted Higgs production from gluon fusion as a complementary approach.\par

To simplify the extraction of the small amount of high-$p_T$ signal from the background, we focus on 
the clean decay of a Higgs to two leptons $\ell=e^\pm,\mu^\pm$ and missing
transverse momentum $\slashed{\mathbf{p}}_T$.  For a \(125\,\text{GeV}\) Standard Model-like Higgs boson, this
occurs almost entirely via $H\to WW^*$ and $H\to \tau\tau$; we will focus on these two channels
separately as detailed in Section~\ref{sec:SM}.\par

The organization of the paper is as follows. In Section~\ref{NewPhysics} we discuss some examples of
beyond the Standard Model physics which motivate this analysis. Section~\ref{MC} outlines how we
generated our signal and background samples.  Section~\ref{sec:SM} contains our signal versus
background analyses for boosted $H\to2\ell+\slashed{\mathbf{p}}_T$ in the Standard Model. Section
\ref{sec:discussion} contains a discussion of the analysis and we conclude in Section~\ref{conc}.

\section{new physics models} \label{NewPhysics}

\subsection{Minimal Composite Higgs Model}
\label{sec:mchm}
In the Minimal Composite Higgs Model (MCHM) \cite{Agashe:2004rs}, electroweak symmetry is broken
dynamically by a strong interaction based on the coset \(SO(5)/SO(4)\). For reviews of MHCM see \cite{Contino:2010rs, Bellazzini:2014yua}.
In this class of models, the
Higgs arises as a pseudo-Nambu-Goldstone Boson (pNGB) of the symmetry breaking which naturally
explains its small mass. Fermionic resonances of the strong sector, coming in multiplets of
\(SO(4)\), will contribute to the gluon fusion loop diagram. These resonances also mix
with the SM fermions and thus modify their couplings to the Higgs. Interestingly, the contributions
of these resonances to the sum of the coefficients \(\kappa_g\) and \(c_t\) cancel exactly in a
broad class of MCHM models and lead, up to small corrections which are negligible at the LHC \cite{Gillioz:2012se}, to
\cite{Falkowski:2007hz,Low:2010mr,Azatov:2011qy,Delaunay:2013iia,Montull:2013mla}
\begin{equation}
  \label{eq:7}
  c_t+\kappa_g=f_g(\xi)
\end{equation}
where \(f_g\) is a function satisfying \(f_g(\xi\to 0)=1\) with \(\xi\equiv v^2/f^2\) and \(f\) is
the decay constant of the non-linear sigma model. The gluon fusion cross-section is therefore
independent of the mass spectrum of the fermionic resonances, and for small \(\xi\) is even
SM-like. This makes it impossible to find traces of the top partner spectrum in the inclusive gluon fusion
process.\par
While the resonances are needed to cut off the UV-divergences of the Higgs mass and thus must not be
too heavy to avoid excessive fine tuning, they should still be heavy
enough to allow for an effective description of the boosted Higgs production. In
\cite{Grojean:2013nya} it was shown that as long as the mass of the lightest resonance is at least
of the order of the Higgs transverse momentum, the result of the calculation in the heavy top limit
lies within \(\mathcal{O}(10\%)\) of the full calculation. Considering that the masses
of the resonances have to be heavier than $600-800~{\rm GeV}$ depending on the representation~\cite{Chatrchyan:2013uxa,ATLAS:2013ima,TheATLAScollaboration:2013oha,TheATLAScollaboration:2013jha,deSimone:2012fs,Aguilar-Saavedra:2013qpa,Buchkremer:2013bha,Azatov:2013hya,Kearney:2013oia,Delaunay:2013pwa},
the effective description is well justified within the scope of the paper.

\subsection{Supersymmetry}
\label{sec:SUSY}
In the minimal supersymmetric SM (MSSM), an analogous flat direction of the inclusive cross-section exists which can be 
resolved by looking at boosted Higgs shapes. 
For certain choices of the stop masses \(m_{\tilde{t}_{1}}\), \(m_{\tilde{t}_{2}}\) and \(A_t\), the
effects of two contributions cancel, yielding a SM-like inclusive signal strength
\cite{Kunszt:1991qe,Barger:1991ed,Baer:1991yc,Gunion:1991er,Gunion:1991cw, Djouadi:1991tka,Spira:1995rr,Dawson:1990zj,Graudenz:1992pv}. 
Assuming the MSSM is in the decoupling limit, and neglecting small D-term contributions, the inclusive
signal strength is given by \cite{Haber:1984zu}
\begin{equation}
  \label{eq:5}
  \frac{\Gamma(gg\to H)}{\Gamma(gg\to H)_{SM}}=\left(1+\Delta_t\right)^2
\end{equation}
where
\begin{equation}
  \label{eq:6}
  \Delta_t\approx\frac{m_t^2}{4}\left(\frac{1}{m^2_{\tilde{t}_1}}
    +\frac{1}{m^2_{\tilde{t}_2}}-\frac{(A_t-\mu/\tan\,\beta)^2}{m^2_{\tilde{t}_1}m^2_{\tilde{t}_2}}\right)
\end{equation}
quantifies the deviation from the SM value and can vanish due to the relative minus sign. A $125~\mathrm{GeV}$
Higgs can easily achieved by extending the MSSM by additional D- or F-terms which should, of course,
not have a major impact on the couplings of the SM-like lightest Higgs.\par
Since the \(A_t\)-dependent parts of the production cross-section are less sensitive to the boost of
the Higgs than the \(A_t\)-independent ones, the aforementioned degeneracy gets broken
in the boosted regime. Therefore the non-SM nature of the Higgs
production can be revealed by looking at the boosted production. Moreover this can make light stops~\cite{Aad:2013ija,
  TheATLAScollaboration:2013aia, TheATLAScollaboration:2013xha, TheATLAScollaboration:2013tha,
  TheATLAScollaboration:2013gha, ATLAS:2013pla, ATLAS:2013bma, ATLAS:2013cma,Chatrchyan:2013xna, CMS:2013ffa, CMS:2013ida, CMS:zrk}\cite{Papucci:2014rja,Papucci:2011wy}
accessible which are hidden in the stealth region and challenging to extract given the similarity to the top background
\cite{Delgado:2012eu, Kats:2011it, Fan:2012jf, Fan:2011yu, Bai:2013xla}.  An outline
showing this sensitivity and taking vacuum stability constraints into account has been presented in
\cite{Grojean:2013nya}.

\subsection{Effective description}
\label{sec:effect-descr}
It is useful to parametrize our ignorance of new physics in terms of an effective Lagrangian. Out of the 59
dimension six operators one can add to the SM \cite{Buchmuller:1985jz, Grzadkowski:2010es}, only four can affect the Higgs
production through gluon fusion \cite{Giudice:2007fh, Contino:2013kra, Elias-Miro:2013mua}. These
four operators as well as the other dimension six operators involving the Higgs are already
constrained to some extent by LHC data~\cite{Elias-Miro:2013mua, Pomarol:2013zra, Englert:2014uua,
  Bramante:2014gda,Ellis:2014dva,Dumont:2013wma}. 
We will focus on CP-conserving effects and omit the
CP-violating operator containing the dual of the QCD gauge field strength. The
remaining three important operators are
\begin{equation}
  \label{eq:2}
  \mathcal{O}_y=\frac{y_t}{v^2}\left|H\right|^2\bar{Q}_L \widetilde{H} t_R,\quad
  \mathcal{O}_H=\frac{1}{2v^2}\partial_\mu\left|H\right|^2\partial^\mu\left|H\right|^2,\quad \text{and}\quad
  \mathcal{O}_g=\frac{\alpha_S}{12 \pi v^2}\left|H\right|^2 G_{\mu\nu}^a G^{a\,\mu\mu}.
\end{equation}
After adding them to the SM Lagrangian and extracting the terms relevant for the gluon fusion
process we obtain
\begin{equation}
  \label{eq:3}
  \mathcal{L}_{\text{eff}}=-c_t
  \frac{m_t}{v}\bar{t}tH+\kappa_g\frac{\alpha_S}{12\pi}\frac{h}{v}G^a_{\mu\nu}G^{a\,\mu\nu} + \mathcal{L}_{\text{QCD}},
\end{equation}
where \(c_t=1-\text{Re}(C_y)-C_H/2\) scales the top Yukawa coupling which enters the process via the
top loop and \(\kappa_g=C_g\) controls the direct gluon-Higgs interaction. The \(C_i\) are the
coefficients of the corresponding operators in \eqref{eq:2} and the coefficients are chosen such
that for \(c_t=1\) and \(\kappa_g=0\) the SM Lagrangian is obtained.\par
The full matrix element for boosted Higgs production is then given by\footnote{In the SM the effects
  of the bottom loop are within a few percent if the boost of the Higgs exceeds
  \(\mathcal{O}(50\,\text{GeV})\) \cite{Bagnaschi:2011tu, Mantler:2012bj, Grazzini:2013mca} and are therefore neglected.}
\begin{equation}
  \label{eq:1}
  \mathcal{M}(c_t,\kappa_g)=c_t \mathcal{M}_{IR}+\kappa_g \mathcal{M}_{UV}
\end{equation}
where \(\mathcal{M}_{IR}\) is the matrix element taking the full top mass dependence into account
\cite{Baur:1989cm} and \(\mathcal{M}_{UV}\) is the one obtained from \(\mathcal{M}_{IR}\) in the
heavy top limit or equivalently from the tree-level diagram generated by \(\mathcal{O}_g\).  From
Eq. \eqref{eq:1} we see that the differential cross-section, normalized by the SM value, can be
described as
\begin{equation}
  \label{eq:semi-numerical-formula}
  \frac{\sigma({p_T^{\rm cut}})}{\sigma^{SM}({p_T^{\rm cut}})}
  = \frac{\int_{p_T^{\rm cut}}^{\infty} d p_T d \Omega | c_t {\cal M}_{IR}(m_t) + 
    \kappa_g {\cal M}_{UV}|^2}{\int_{p_T^{\rm cut}}^{\infty} d p_Td \Omega |{\cal M}_{IR}(m_t)|^2}
  = (c_t +  \kappa_g)^2  + \delta(p_T^{\rm cut}) c_t \kappa_g  + \epsilon(p_T^{\rm cut}) \kappa_g^2, 
\end{equation}
where

\begin{eqnarray}
  &&\delta(p_T^{\rm cut}) = 
  \frac{2\int_{p_T^{\rm cut}}^{\infty} d p_T d \Omega  Re({\cal M}_{IR}(m_t) {\cal M}_{UV}^*)}{\int_{p_T^{\rm cut}}^{\infty} d p_T d \Omega  |{\cal M}_{IR}(m_t)|^2} - 2,\\
  &&\epsilon(p_T^{\rm cut}) = \frac{\int_{p_T^{\rm cut}}^{\infty} d p_Td \Omega   |{\cal M}_{UV}|^2}{\int_{p_T^{\rm cut}}^{\infty} d p_T d \Omega  |{\cal M}_{IR}(m_t)|^2} - 1.
\end{eqnarray}

For small $p_T^{\rm cut}$, the coefficients $\delta, \epsilon$ are very small, modifying the cross
section only by a few percent, which is less than the uncertainty expected in the inclusive Higgs cross
section measurements~\cite{ATL-PHYS-PUB-2013-014,Liss:1564937,CMS-NOTE-2012-006}.
This is what is expected due to the very good description of both
the top and the new particle loop by the effective interaction. On the other hand, $\delta,
\epsilon$ grow significantly as $p_T^{\rm cut}$ increases, and they become ${\cal O}(1)$ for
$p_T^{\rm cut}> 300$ GeV~\cite{Grojean:2013nya}.  It means we can break the degeneracy by measuring
the Higgs $p_T$ distribution while we cannot break the degeneracy along $c_t + \kappa_g = const.$
direction only by determining the inclusive cross-section.

\section{Event generation} \label{MC}

\subsection{Signal sample}
\label{sec:signal-sample}

In this paper we consider $H+$jet events with subsequent $H$ decays to $WW^* \to \ell^+\ell^- \nu
\bar{\nu}$ and $\tau^+\tau^-$ modes as a signal.  The signal events are generated with
\texttt{MadGraph5}, version 1.5.15~\cite{Alwall:2011uj} and showered with
\texttt{HERWIG++}~\cite{Bahr:2008pv, Arnold:2012fq, Bellm:2013lba}, where only $WW^*$ and
$\tau^+\tau^-$ decays are specified.\par
We have used \texttt{MadGraph5} to generate \(H+\)jet events using the `HEFT' model with SM couplings which
makes use of the low energy theorem. The generated cross-section is proportional to
\(\left|\mathcal{M}(0,1)\right|^2\) and does not take into account finite top mass effects which are
crucial to our analysis. To obtain the correct weight of the events we reweighted them by a weight
factor
\begin{equation}
  \label{eq:4}
  w(c_t,\kappa_g)=\frac{\left|\mathcal{M}(c_t,\kappa_g)\right|^2}{\left|\mathcal{M}(0,1)\right|^2}
\end{equation}
making use of our own code, which is based on an implementation of the formulas for the matrix
elements given in \cite{Baur:1989cm} and also calculated in \cite{Ellis:1987xu}.
At present no finite top mass NLO computation of the SM Higgs $p_{T}$ spectrum is available.  
An exact NLO prediction of SM Higgs $p_{T}$ spectrum
would be very desirable and help to exploit the full potential of this observable.
Recent progress in the precision prediction of $h+\mathrm{jet}$ can be found in Refs.~\cite{Harlander:2012hf,Boughezal:2013uia,Hoeche:2014lxa}.
We will approximate the NNLO (+ NNLL) result of 49.85 pb~\cite{Dittmaier:2011ti,Anastasiou:2008tj,deFlorian:2009hc,Bonvini:2014joa}
by multiplying the exact LO result with a $K$ factor of 1.71.

\begin{figure}[b]
\begin{center}
  \includegraphics[width=0.325\textwidth]{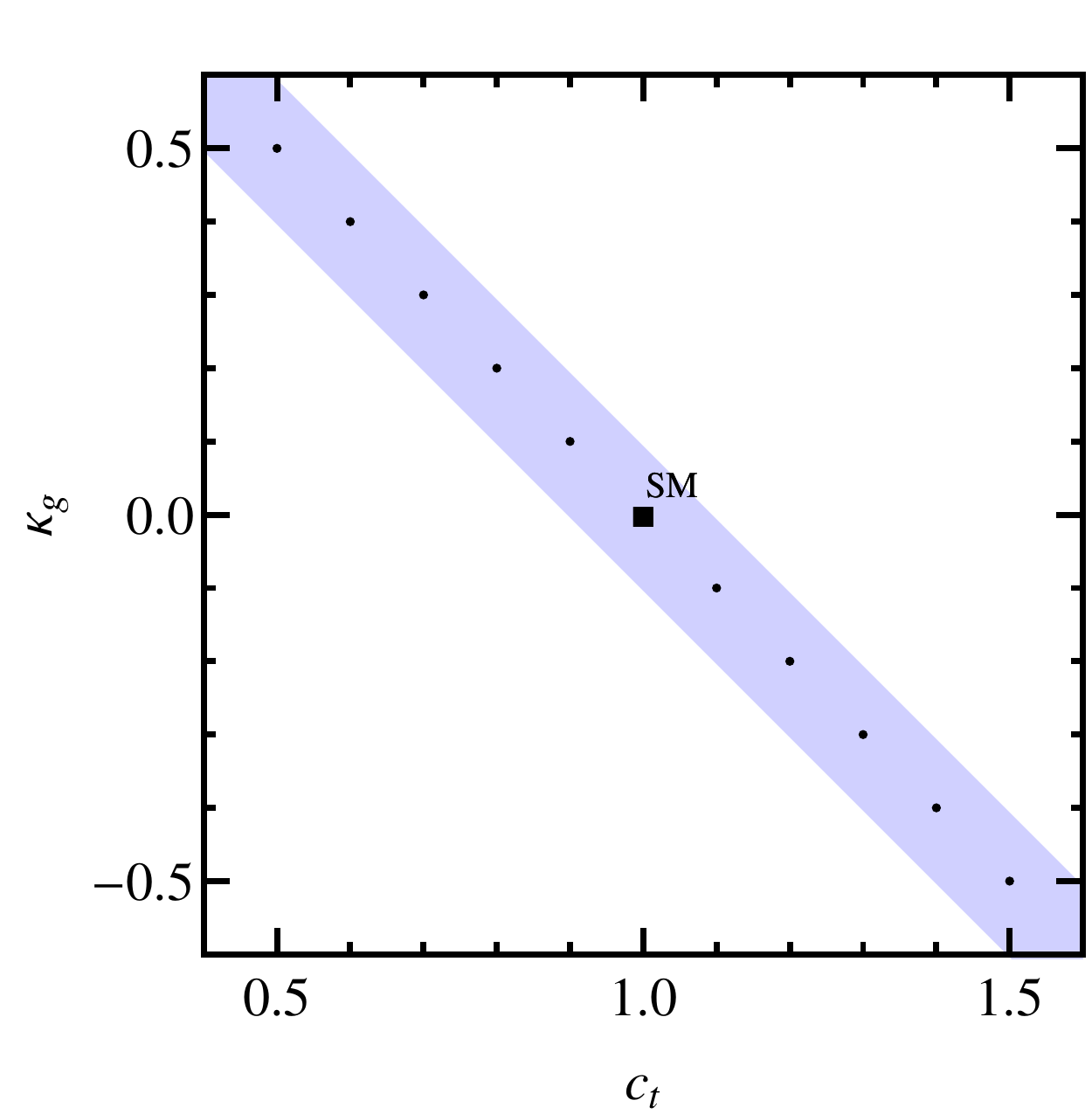}\qquad\qquad\qquad
  \includegraphics[width=0.32\textwidth]{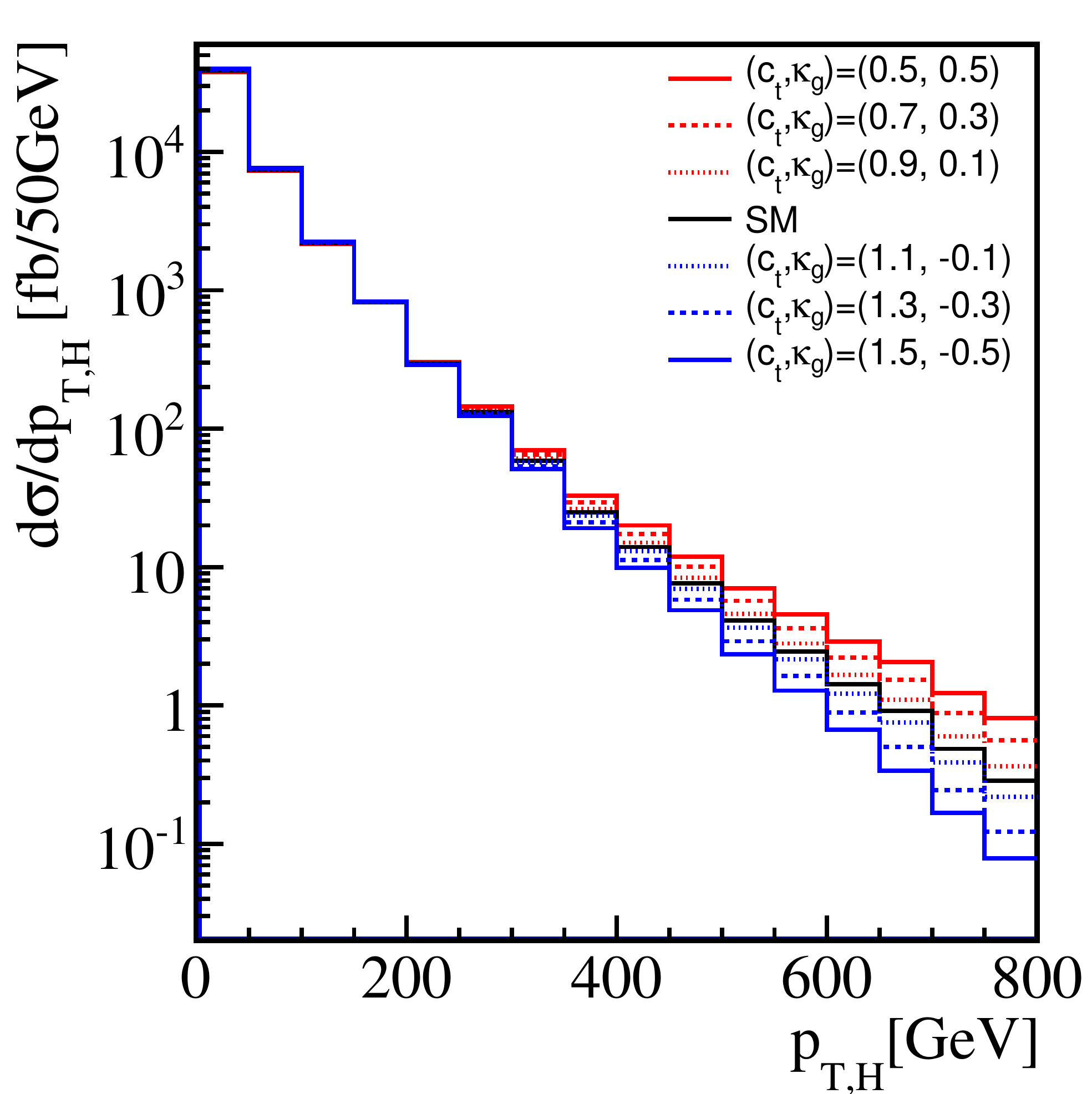}
  \caption{ Left panel: model points generated for this analysis in $(c_t, \kappa_g)$ plane. The shaded area shows
  parameter space which gives the inclusive cross-section consistent to the SM prediction within 20\%.
  Right panel: parton level $p_{T,H}$ distributions for the SM, and $(c_t,\kappa_g)=(1 -\kappa_g,
    \kappa_g)$ with $\kappa_g =\pm 0.1,\pm 0.3,\pm 0.5$.  }
  \label{fig:NPmodels}
  \end{center}
\end{figure}

We reweight the events for points along the line $c_t + \kappa_g = 1$ for $\kappa_g\in\left[ -0.5,
  0.5\right]$ with steps of 0.1, as shown in the left panel of Fig.~\ref{fig:NPmodels}. This is
consistent with the SM inclusive Higgs production cross-section.  
The size of $c_t$ alone is only weakly constrained by the current $t\bar{t}H$
measurement.  Although we only consider the most difficult points satisfying $c_t + \kappa_g = 1$ (i.e. an 
exactly SM-like inclusive cross-section), an analysis
along different $c_t + \kappa_g = const.$ lines would be straightforward as a different choice 
essentially just corresponds to an overall rescaling of the signal. 

The right panel of the Fig.~\ref{fig:NPmodels} shows the $p_{T,H}$ distributions for several
model points. In the region with low $p_{T,H}$, the distributions are degenerate but for high $p_{T,H}$ the
distributions start to split. For the model points with $\kappa_g>0$ we see an enhancement in the high
$p_{T,H}$ region while we see the suppression for the model points with $\kappa_g<0$.
Table~\ref{NPpartonlevel} shows the Higgs production cross-sections relative to the SM value for
several model points \((c_t,\kappa_g)\) and $p_{T,H}$ cuts.  As one can see, for \(p_{T,H}>10\,\text{GeV}\) the cross-sections are essentially the same as the SM value within 3\%, while
for increasing $p_T^{\rm cut}$, significant differences from the SM predictions can be observed.  For
the model point $(c_t,\kappa_g)=(0.7,0.3)$, for example, a 6\% difference would be observed for
$\sigma(p_{T,H} > 200~\mathrm{GeV}$), and a $\sim$ 20\% difference for $\sigma(p_{T,H} > 300~\mathrm{GeV}$). We will see that 
these effects are comparable to the sensitivity of the boosted Higgs shape measurements, see Section \ref{sec:SM}. For very hard cuts, ${\cal O}(1)$ differences can be
observed, as can be seen from the cross-section ratios for $p_{T,H} > 500$~GeV and harder.

\begin{table}[h!]
\begin{center}
\renewcommand{\arraystretch}{1.4}
  \begin{tabular}{l|ccccccc}
     & (0.5,0.5) & (0.7,0.3) & (0.9,0.1) & {\bf SM} & (1.1,-0.1) &(1.3,-0.3) &(1.5,-0.5)  \cr
    \hline
    $p_{T,H}>{}10$~GeV & 0.9733&  0.9838&  0.9945&  1.0000&  1.0055&  1.0167&  1.0281\cr 
    \hline
    $p_{T,H}>100$~GeV & 1.0044&  1.0012&  0.9999&  1.0000&  1.0006&  1.0031&  1.0076\cr 
    $p_{T,H}>200$~GeV & 1.1166&  1.0646&  1.0198&  1.0000&  0.9820&  0.9513&  0.9277\cr 
    $p_{T,H}>300$~GeV & 1.3450&  1.1921&  1.0591&  1.0000&  0.9459&  0.8526&  0.7791\cr 
    $p_{T,H}>400$~GeV & 1.6531&  1.3590&  1.1087&  1.0000&  0.9023&  0.7397&  0.6210\cr 
    $p_{T,H}>500$~GeV & 2.0233&  1.5520&  1.1633&  1.0000&  0.8573&  0.6340&  0.4932\cr 
    $p_{T,H}>600$~GeV & 2.4869&  1.7871&  1.2274&  1.0000&  0.8076&  0.5279&  0.3882\cr 
    $p_{T,H}>700$~GeV & 3.1213&  2.1003&  1.3093&  1.0000&  0.7482&  0.4172&  0.3161\cr 
    $p_{T,H}>800$~GeV & 3.7427&  2.3989&  1.3841&  1.0000&  0.6981&  0.3411&  0.3129\cr 
    \hline
  \end{tabular}
  \caption{Cross-sections normalized by the SM value after applying several $p_{T,H}$ cuts in parton level for several model points $(c_t,\kappa_g)$.}
  \label{NPpartonlevel}
  \end{center}
\end{table}


\subsection{Background sample}
\label{sec:background-sample}

We include $W$+jets, $Z$+jets and $t\bar{t}$+jets as  background processes which we have generated
with \texttt{ALPGEN} + \texttt{PYTHIA}~\cite{Mangano:2002ea,Sjostrand:2006za}. Since we consider
{\it boosted} Higgs reconstruction and since we will require the existence of one hard recoil jet, we
apply a pre-selection cut in the generation step, where we demand at least one recoil parton of
$p_{T}>150$~GeV.  We merge up to two partons for $WW$+jets and $Z$+jets, and up to one parton for
$t\bar{t}$+jets using the MLM matching scheme~\cite{Mangano:2001xp,Hoche:2006ph}.  As we only consider
the dilepton mode in this paper we preselect the $W$ decay mode, including $W$ from tops only with
leptons, $e, \mu$, and $\tau$.  For the $Z$ decay, we consider only $Z \to \tau^+\tau^-$ since for the
other leptonic decay modes we can reconstruct the $Z$-peak and reject them.  We rescale
the $t\bar{t}$ sample to obtain a NLO inclusive cross
section of 918 pb~\cite{Nason:1987xz,Beenakker:1988bq,Moch:2008qy}. For the $Z+$jets
and $WW+$jets samples we used LO cross-sections.

Our analysis is performed at particle level with a simple detector simulation with the granularity
resolution of $\Delta \eta \times \Delta \phi=0.1 \times 0.1$.  After removing the isolated leptons,
the energy of the remaining visible particles falling into each cell are summed up.  Cells with
transverse energy above $0.5~\mathrm{GeV}$ are used for the further jet reconstruction.  

Jet clustering was performed using the \texttt{FastJet}~\cite{Cacciari:2011ma} version
3.0.4.  We use the Cambridge-Aachen (C/A) algorithm~\cite{Dokshitzer:1997in,Wobisch:1998wt} with $R=0.5$ 
for normal jet and $b$-tag jet definition. We also define `fat' jets, as explained later, defined using the
C/A algorithm with $R=1.5$.

In this paper, we only consider the events with isolated leptons for simplicity. There is room for
improving the analysis with hadronic tau modes with tau tagging for example \cite{Gripaios:2012th,Katz:2010iq}, which is,
however, beyond the scope of our current study.

\section{Boosted $H\to2\ell+\slashed{\mathbf{p}}_T$ in the Standard Model} \label{sec:SM}

In our notation a subscript $\ell$ will denote \textit{leptonically-decaying}: $\tau_\ell$ thus
represents $\tau\to \ell+2\nu$, $W_\ell$ is mostly $W\to l\nu$ with some $W\to \tau_\ell\nu_\tau$,
and $t_\ell$ is $t\to bW_\ell$.  The decay $H\to2\ell+\slashed{\mathbf{p}}_T$ is mostly\footnote{
  The branching ratios for $H\to W_\ell W_\ell^*$ and $H\to \tau_\ell\tau_\ell$ are $BR=1.4\%$ and
  $0.77\%$ respectively; $H\to ZZ^*\to2\ell+2\nu$ is negligible with $BR=8\times10^{-4}$.} through
$H\to W_\ell W_\ell^*$ and $H\to \tau_\ell\tau_\ell$.  As noted
in~\cite{Dittmar:1997nea,Dittmar:1996ss}, in the decay $H\to WW^*\to 2\ell+2\nu$ spin correlations
ensure that the two lepton momenta have similar directions, as do the two neutrino momenta.  In
$H\to \tau_\ell\tau_\ell$ however, the two $\tau$ leptons are back-to-back in the Higgs rest frame,
and each of them gives rise to a highly collimated $\ell+2\nu$ trio.  These two facts imply that
for a boosted $H\to2\ell+\slashed{\mathbf{p}}_T$ decay, the $\slashed{\mathbf{p}}_T$ is typically
outside the lepton pair for the $H\to W_\ell W_\ell^*$ contribution and inside the lepton pair for
$H\to \tau_\ell\tau_\ell$, as shown in Table~\ref{tab:leptonMETpositions}.  We use
this binary criterion -- $\slashed{\mathbf{p}}_T$ inside or outside the leptons -- to split our
analysis into two sub-analyses, which differ in their background compositions as well as signals.

\begin{table}[ht]
  \centering
  \begin{tabular}{l | *{1}{m{0.25\textwidth}} c *{1}{m{0.25\textwidth}} }
    \vphantom{$\sum_g$} &\begin{center}$H\to W_\ell W_\ell^*$\end{center} &&\begin{center} $H\to \tau_\ell\tau_\ell$\end{center} \\
    \hline
    Higgs rest frame  \hspace*{5mm} &
    \begin{center}\includegraphics[width=\linewidth]{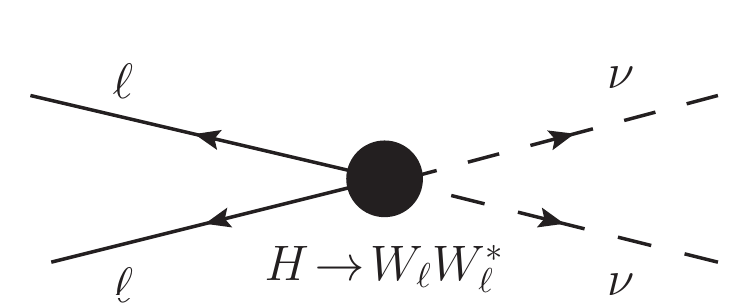}\end{center} &\hphantom{blah}&
    \begin{center}\includegraphics[width=1.3\linewidth]{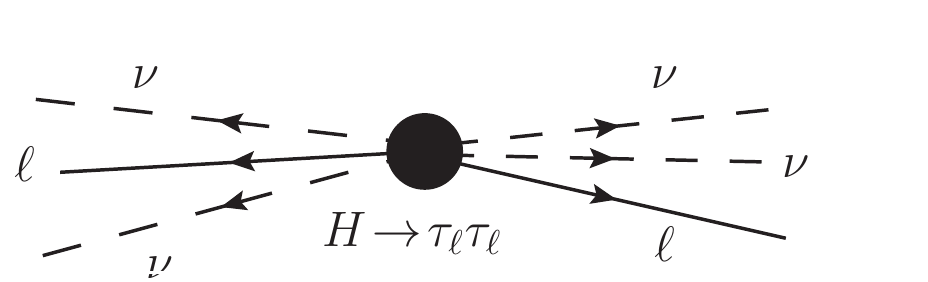}\end{center}
    \\
    boosted Higgs &
    \begin{center}\includegraphics[width=0.7\linewidth]{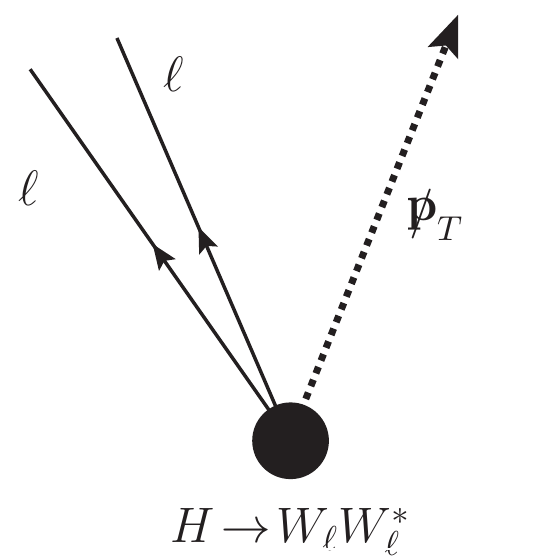}\end{center} &&
    \begin{flushright}\includegraphics[width=0.6\linewidth]{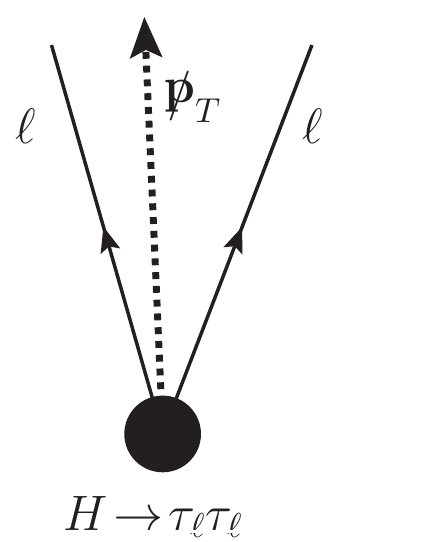}\end{flushright}
  \end{tabular}
  \caption{Showing the difference in the relative positioning of the neutrinos/$\slashed{\mathbf{p}}_T$ and the dilepton system between $H\to W_\ell W_\ell^*$ and $H\to \tau_\ell\tau_\ell$ decays.}
  \label{tab:leptonMETpositions}
\end{table}

\subsection{Common Cuts for $H\to \tau_\ell\tau_\ell$ and $H\to W_\ell W_\ell^*$} \label{sharedCuts}

In both of our sub-analyses the cuts begin by requiring the following:
\begin{itemize}
\item Two opposite-sign isolated leptons each having \mbox{$p_{T} > 10$~GeV} and $|\eta|<2.5$. 
         If a third isolated lepton with \mbox{$p_{T} > 5$~GeV} and $|\eta|<2.5$ is present, the event is vetoed.
   Our isolation criterion is $E_{T,{\rm had}}^{R<0.2}/E_{T,\ell} < 0.1$, where $E_{T,
    {\rm had}}^{R<0.2}$ is the sum of transverse energies over all hadronic activity in the cone
  $\Delta R<0.2$ around the lepton.  
  (The signal leptons are typically hard, so our $p_T$ threshold could be raised with minimal loss of
  efficiency.)  
\item A dilepton mass $m_{\ell\ell}$ exceeding 20~GeV, which is necessary in practice to suppress Drell-Yan
  dilepton production (not simulated here).
\item At least 200~GeV of transverse momentum for the system obtained by vectorially summing the
  dilepton and missing transverse momenta:
  \begin{equation} \label{eq:ptH} |{\mathbf p}_{T,H}| \equiv |\slashed{\mathbf{p}}_T + {\mathbf
      p}_{T,\ell_1} + {\mathbf p}_{T,\ell_2}| > 200~{\rm GeV}.
  \end{equation}
  The system thus defined has a transverse (but not longitudinal) momentum coinciding with the Higgs
  in the case of the signal: herein lies our restricted focus on highly energetic/boosted
  Higgs bosons.
\item One `fat' jet, resulting from clustering using the C/A algorithm with a distance
  parameter $R_{\rm jet}= 1.5$.  This jet should be very hard:
  \begin{equation}
    p_{T,j}> 200~{\rm GeV}.
  \end{equation}
  The presence of a very hard jet coincides with our parton-level picture of the signal process: a
  boosted Higgs recoiling against a gluon/quark.  Defining geometrically large `fat' jets allows us
  to capture the radiation emitted by this gluon/quark (which might otherwise be clustered into a
  separate jet when clustering with traditional `skinny' jets).  We veto if there is a second fat
  jet with $p_{T,j}>100$~GeV.  Vetoing on additional hadronic activity beyond the first hard fat jet
  suppresses higher-multiplicity backgrounds, i.e. $t\bar{t}+$jets. Not vetoing additional fat jets approximately doubles the $t \bar{t}$ background, while the signal increases by roughly $30 \%$. These numbers even hold in case regular jets with a cone size of $R=0.4$ are vetoed instead. When vetoing jets large logarithms $\sim \ln^2 (\sqrt{\hat{s}}/p_{T,\mathrm{veto}})$ can be induced which need to be resummed \cite{Berger:2010xi,Banfi:2012jm}. However, due to the high veto scale we do not expect these contributions to spoil the reliability of our analysis. As an alternative to jet vetos, 2-jet observables can be used to disentangle signal from background in this process \cite{Buschmann:2014twa, Dolan:2014upa}. 
\item Zero $b$-tags.  This considerably reduces the (until now dominant) $t\bar{t}+$jets background
  while having negligible effect on the signal.  We re-cluster the hadronic activity into jets,
  again using the C/A algorithm but now with $R_{\rm jet}= 0.5$, to use for the
  $b$-tagging.  We assume a flat 70\% (1\%) efficiency for $b$ (light quark or gluon) initiated jets,
  i.e. a 30\% (99\%) probability for such a jet {\it not} to provoke the veto.  We only consider
  $b$-jets of $p_{T,b}>30$~GeV and $|\eta_b| < 2.5$.
\end{itemize}

The efficiencies of these cuts for the signal and various backgrounds are shown in the first part of
Table~\ref{tab:htautau}.  At this stage the backgrounds from $WW$/$Z$/$t\bar{t}$ +jets are seen to
contribute at similar levels.  The set of cuts described so far are common to both our $H \to
\tau_\ell \tau_\ell$ and $H \to W_\ell W_\ell^*$ analyses; from this point onwards they diverge.

\subsection{$H \to \tau_\ell \tau_\ell$ analysis} 
\label{TauTau}

The Higgs mass in the decay $H\to\tau\tau$ can be reconstructed using the {\it collinear
approximation}~\cite{Ellis:1987xu}. The large hierarchy between the
Higgs and tau masses ensures a very large boost for the taus, highly collimating their
visible and invisible decay products. We can approximate the neutrino
momenta by a decomposition of the missing transverse momentum, which assumes that 
each invisible momentum is parallel to the corresponding visible momentum.
(This procedure  can be extended to decays of more than one particle, see~\cite{Spannowsky:2013qb}). 
As was noted in~\cite{Ellis:1987xu}, and further explored in~\cite{Mellado200560}, this procedure gains sensitivity with
increasing transverse momentum of the Higgs -- i.e. when the Higgs recoils against a hard jet.  
It suffers for a low-$p_T$ Higgs because the two $\tau$ daughters are then nearly back-to-back,
providing a poor basis for the $\slashed{\mathbf{p}}_T$ decomposition.  For our high-$p_T$ Higgs study
the mass reconstruction of the signal in this manner is very good and provides a sharp peak\footnote{We have also tried a
  reconstruction using the $m_{T,\text{Bound}}$-if-not-$m_{T,\text{True}}$ prescription
  of~\cite{Barr:2011he} and found very similar results to the collinear approximation; the former is
  expected to be preferred at lower boosts of the Higgs which we do not consider here.}.

In more detail, the Higgs mass in $H \to \tau_\ell \tau_\ell$ is reconstructed via the collinear
approximation as follows.  We require the missing transverse momentum $\slashed{\mathbf{p}}_T$ to be inside
the two leptons (more precisely, projecting the two lepton momenta into the transverse plane defines two segments; `inside'   
the leptons means inside the smaller segment). We decompose $\slashed{\mathbf{p}}_T$ as a linear combination of the 
two lepton momenta (defining for it a longitudinal component in the process):
\begin{equation}
  \slashed{\mathbf{p}}_T = {\mathbf p}_{T,\nu_1,{\rm col}} +  {\mathbf p}_{T,\nu_2,{\rm col}}: \qquad
  {\mathbf p}_{\nu_1,{\rm col}} = \alpha_1  {\mathbf p}_{\ell_1}, \ \ \ 
  {\mathbf p}_{\nu_2,{\rm col}} = \alpha_2  {\mathbf p}_{\ell_2}.
\end{equation}
The requirement that $\slashed{\mathbf{p}}_T$ be inside the leptons is 
equivalent to demanding that the decomposition coefficients are both positive:
\begin{equation}
  \alpha_1  > 0 \qquad {\rm and} \qquad \alpha_2  > 0.
\end{equation}
${\mathbf p}_{\nu_1,{\rm col}}$ and ${\mathbf p}_{\nu_2,{\rm col}}$ thus defined approximate the
neutrino three-momenta.  Promoting them to massless four-momenta and adding them to the lepton
four-momenta gives an approximate Higgs four-momentum, the mass of which we refer to as the collinear Higgs mass:
\begin{equation}
  p_{\rm col}=p_{\nu_1,{\rm col}} + p_{\nu_2,{\rm col}} +   {p}_{\ell_1} + {p}_{\ell_2}, \qquad  M_{\rm col}^2= p_{\rm col}^2.
\end{equation}

\begin{figure}[t]
  \includegraphics[width=0.32\textwidth]{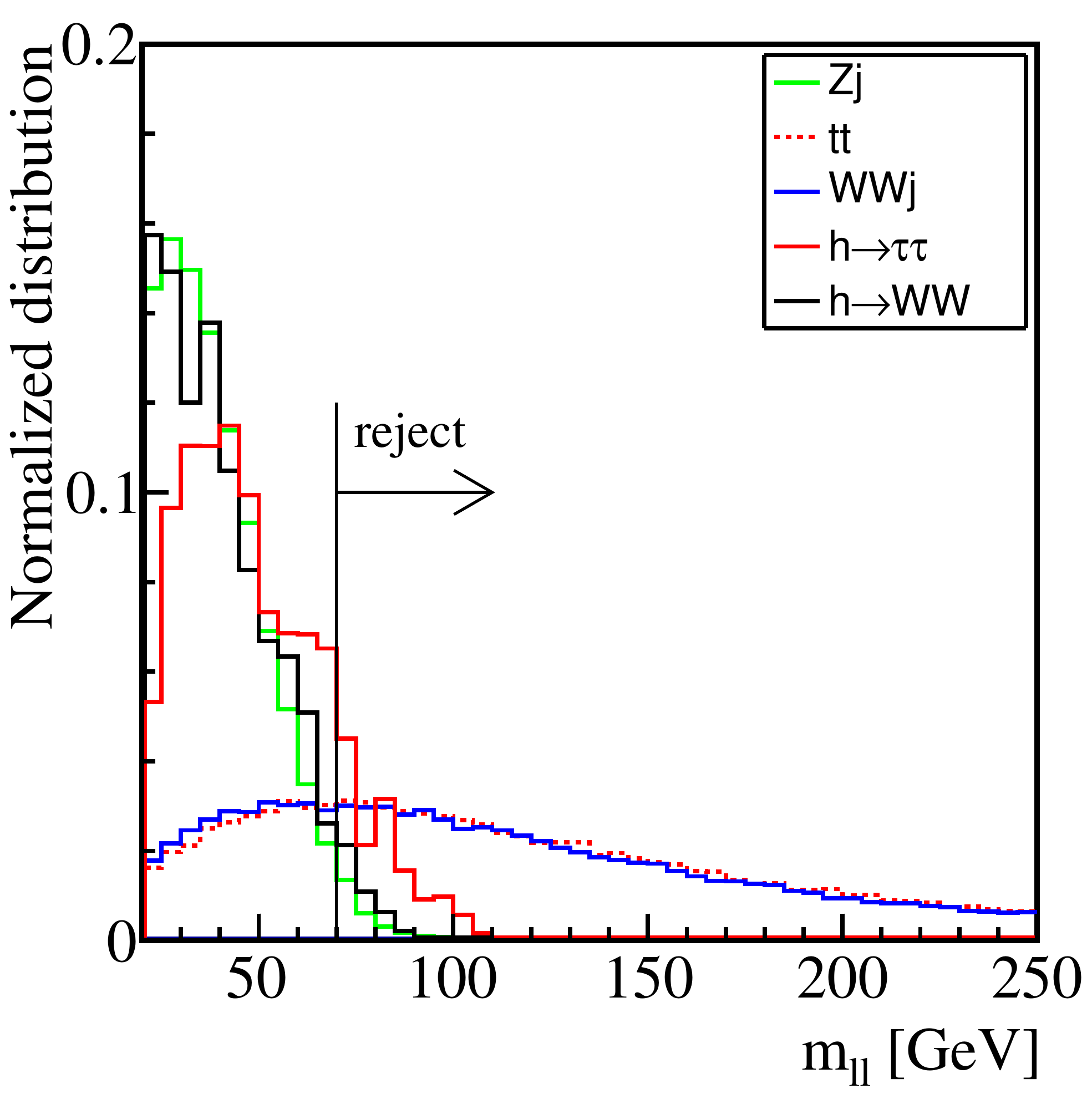}
  \includegraphics[width=0.32\textwidth]{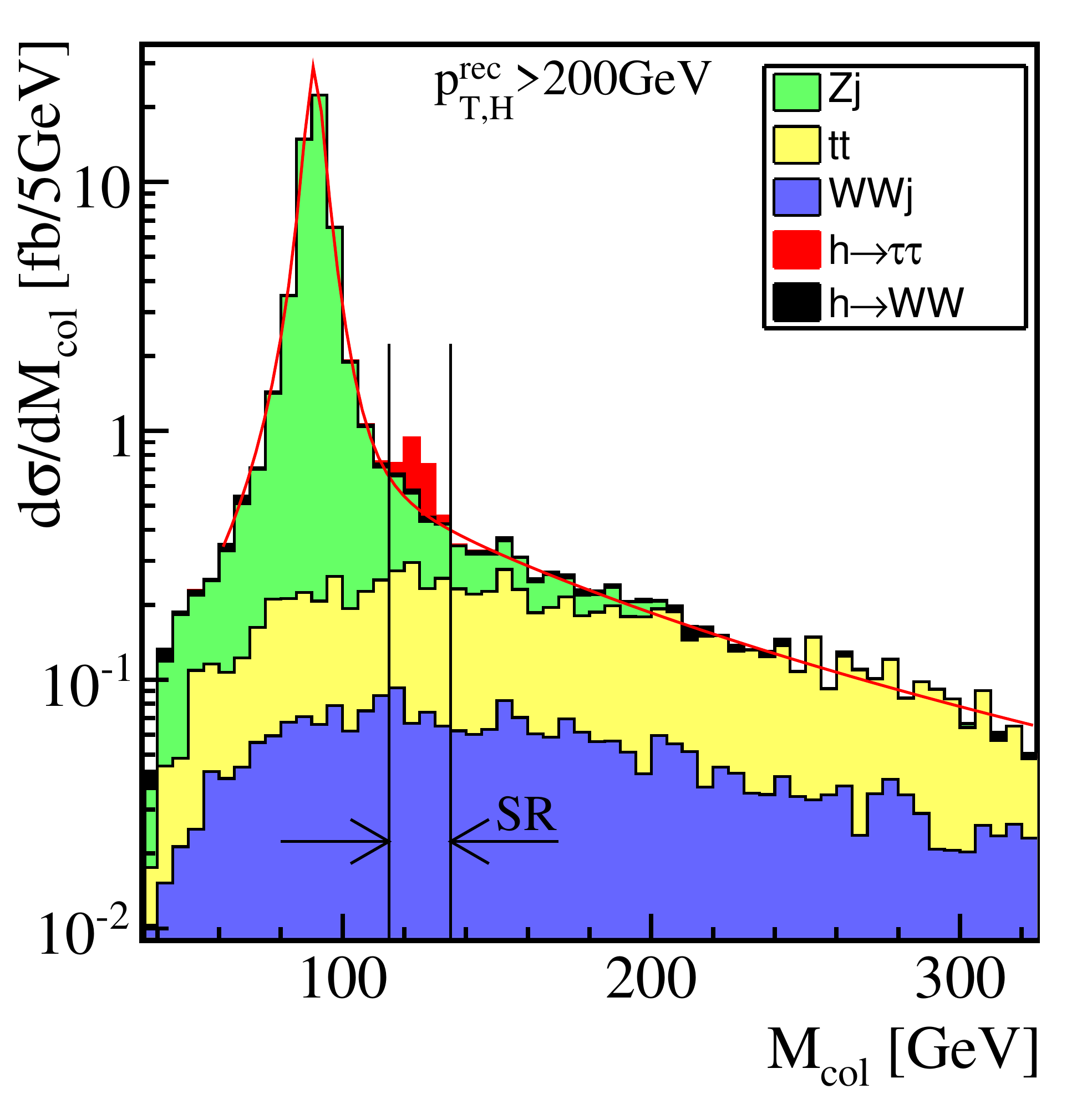}
  \includegraphics[width=0.32\textwidth]{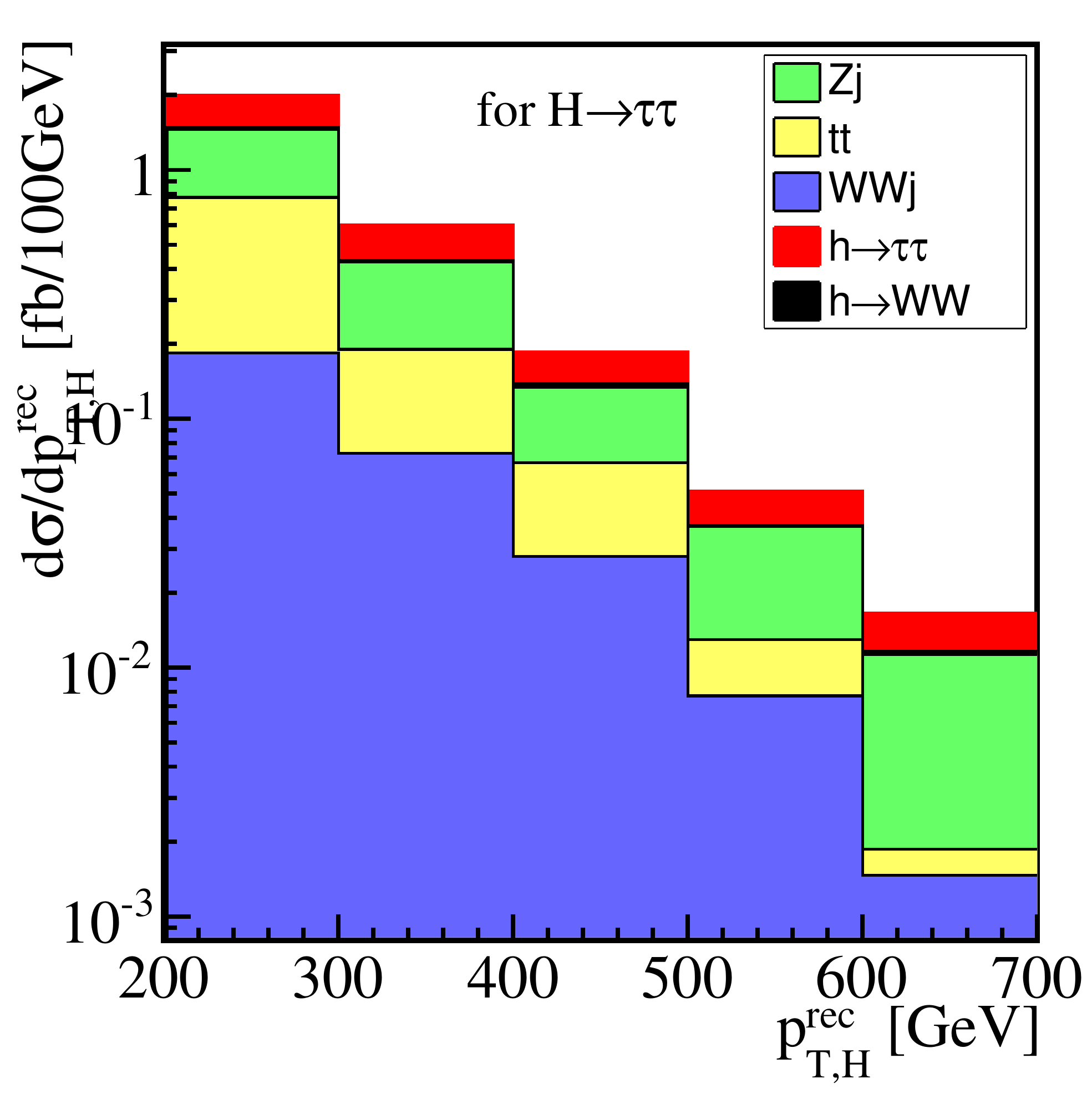}
\caption{
Left panel: the invariant mass of the two leptons, $m_{\ell \ell}$, after cut 6.
Central panel: The collinear mass $M_{\rm col}$ after cut 7, stacking the different processes.
Histograms are normalized to the respective cross-sections.
Right panel: stacked distributions of the `Higgs' transverse momentum $p_{T,H}$ (defined in
Eq.~\eqref{eq:ptH}) after selection cut~8, with a logarithmic scale.
}
\label{fig:htautau}
\end{figure}

We apply one more cut before making use of the collinear mass variable: an upper
limit for the dilepton mass, $m_{\ell \ell} < 70$~GeV.  This cut reduces the
$t\bar{t}+$jets and $WW+$jets backgrounds very efficiently while leaving most the $H+$jets signal and $Z\to
\tau\tau$ background (see Fig.~\ref{fig:htautau}, left panel).  At this stage $Z\to \tau\tau$ becomes the dominant background for extracting the
$H \to \tau\tau$ signal.  The size of the $t\bar{t}$ and $WW$ backgrounds can be estimated in a
data-driven way by removing $m_{\ell \ell} < 70$~GeV cuts. We discuss this in detail in Appendix~\ref{appendix}.

The collinear mass is shown in the central panel of Fig.~\ref{fig:htautau}.  Note that
any particle decaying to $\tau_\ell \tau_\ell$ with enough boost that the two $\tau$ are not
back-to-back will have its mass reconstructed by this process; indeed the most striking feature of
the collinear mass distribution is the $Z$ mass peak from the large irreducible $Z \to
\tau_\ell \tau_\ell$ background. A peak due to the signal is visible at $M_{\rm col}\sim m_H=125$~GeV.
By selecting events in the window $|M_{\rm col} - m_H| < 10$~GeV we achieve a $S/B \sim
0.4$ with $S/\sqrt{B} > 9$ for 300~fb$^{-1}$.  The signal is taken to include the $H \to W W^*$
contribution, which contributes about $\sim\!10$\% the $H \to \tau \tau$ selection.  
We estimate the statistical error of the high $p_T$ cross-section measurement with $\sqrt{S+B}/S$.
We obtain uncertainties of 12\% for $\sigma(p_{T,H}>200$~GeV), 22\% for $\sigma(p_{T,H}>300$~GeV), and 41\% for
$\sigma(p_{T,H}>400$~GeV), respectively.  Assuming we can achieve the same efficiencies for high-luminosity run of
the LHC (HL-LHC) at 3 ab$^{-1}$, we obtain $\sim 4$\% for
$\sigma(p_{T,H}>200$~GeV), $\sim 7$\% for $\sigma(p_{T,H}>300$~GeV), and $\sim 13$\% for
$\sigma(p_{T,H}>400$~GeV).

As seen in the central panel of Fig.~\ref{fig:htautau} the smooth side-band distribution can be used
for estimating the background contribution.  We show in Appendix~\ref{appendix} that these side-bands are available even 
after hard $p_{T,H}^{\rm rec}$ cuts.  We therefore expect that a data-driven strategy for background estimation will be 
available, and take the statistical errors as a background uncertainty estimate. There will of course be
further systematic uncertainties induced by MC background modeling.

In this analysis we mostly use the recoiling fat jet to remove the $t\bar{t}+$jets
background. It could be beneficial to make use of the
difference between the jet substructure of gluon and quark
jets~\cite{Gallicchio:2011xq,ATLAS:2012zoa, CMS:2013kfa, Larkoski:2013eya,Mukhopadhyay:2014dsa} since the dominant background 
at the last stage is $Z+$jets, which gives a different fraction of gluon and quark jets
than the $H+$jets signal. We leave this for future work.

\begin{table}[t]
\begin{center}
\renewcommand{\arraystretch}{1.3}
  \begin{tabular}{l|cc|ccc||cc}
    Event rate [fb] & $H \to \tau \tau$ & $H \to W W^*$ &
    $W_\ell W_\ell$+jets
    &$Z_{\to\tau\tau}$+jets & $t_\ell\bar{t}_\ell$+jets
    & $S/B$ & $S/\sqrt{B}$ \cr
    \hline
    0. Nominal cross-section & 3149.779 & 10719.207 & 580.000 & 1.01$\cdot 10^4$ & 1.02$\cdot 10^5$ &-- & -- \cr
    1. $n_{\ell}= 2$, opposite-sign & 118.043 & 323.531 & 195.033 & 347.516 & 3.72$\cdot 10^4$& -- & -- \cr
    2. $m_{\ell\ell} > 20$~GeV  & 117.733 & 264.723 & 189.522 & 315.201 & 3.57$\cdot 10^4$ & -- & -- \cr
    3. $p_{T,H}^{\rm rec} > 200$~GeV & 1.987 & 3.834 & 91.273 & 104.434 & 1.28$\cdot 10^3$ & 0.004 & 2.62 \cr
    4. $n_j^{\rm fat}=1$ ($p_{T,j}>200$~GeV)& 0.957 & 1.858 & 50.443 & 58.810 & 395.602 & 0.006 & 2.17 \cr
    5. $n_b=0$ & 0.940 & 1.825 & 48.855 & 57.068 & 105.851 & 0.01 & 3.29 \cr
    \hline
    6. $\slashed{\mathbf{p}}_T$ inside the two leptons    & 0.923 & 0.533 & 20.215 & 55.551 & 44.050 & 0.01 & 2.30 \cr
    7. $m_{\ell\ell} < 70$~GeV & 0.796 & 0.490 & 3.860 & 53.985 & 8.511 & 0.02 & 2.73 \cr
    8. $|M_{\rm col}  - m_H| < 10$~GeV & 0.749 & 0.046 & 0.298 & 1.019 & 0.758 & 0.38 & 9.56 \cr
    \hline
    \ \ \ \ $p_{T,H}^{\rm rec}>300$~GeV  & 0.234 & 0.012 & 0.115 & 0.343 & 0.166 & 0.39 & 5.40 \cr
    \ \ \ \ $p_{T,H}^{\rm rec}>400$~GeV    & 0.068 & 0.006 & 0.042 & 0.106 & 0.049 & 0.38 & 2.88\cr
    \ \ \ \ $p_{T,H}^{\rm rec}>500$~GeV    & 0.021 & 0.001 & 0.014 & 0.038 & 0.010 & 0.36 & 1.55\cr
    \ \ \ \ $p_{T,H}^{\rm rec}>600$~GeV   & 0.008 & 0.001 & 0.006 & 0.014 & 0.005 & 0.32 & 0.89 \cr
    \hline
  \end{tabular}
  \caption{Cut efficiencies for our analysis aimed at $H \to \tau_\ell \tau_\ell$.
    The values for each process are cross-sections in fb. $S/\sqrt{B}$ has been calculated for ${300~{\rm fb}^{-1}}$.
    The $W$ bosons in our $WW$/$t\bar{t}$ +jets backgrounds 
    were forced to decay to $e$, $\mu$, or $\tau$.
  }
  \label{tab:htautau}
  \end{center}
\end{table}

\subsection{$H \to W_\ell W_\ell^*$ analysis} 
\label{WW} 

Our selection criteria for extracting $H \to W_\ell
W_\ell^*$ from the background begin with those described in Section~\ref{sharedCuts}.  In
Section~\ref{TauTau} we required that the $\slashed{\mathbf{p}}_T$ vector be inside the two lepton
momenta, after which the signal was dominated by $H \to \tau_\ell \tau_\ell$ and the background
by $Z_{\to \tau_\ell \tau_\ell}$+jets.  Here we will remove most of the contribution of these
processes by requiring that $\slashed{\mathbf{p}}_T$ be {\it outside} the two lepton momenta.  This
is equivalent to demanding that the $m_{T2}^{\ell \ell}$ variable~\cite{Lester:1999tx} be greater
than zero, as $m_{T2}^{\ell \ell}=0$ when this is not satisfied -- the `trivial
zero'~\cite{Lester:2011nj}.  In fact we go further and impose
\begin{equation}
  m_{T2}^{\ell \ell} > 10~{\rm GeV}.
\end{equation}
This rejects essentially all of the contributions from $H \to \tau \tau$ and $Z\to \tau
  \tau$+jets, which have the same end point close to $m_\tau$.  Allowing for endpoint smearing
we cut a little harder at $10$~GeV instead of $m_\tau$.

\begin{figure}[b]
  \includegraphics[width=0.32\textwidth]{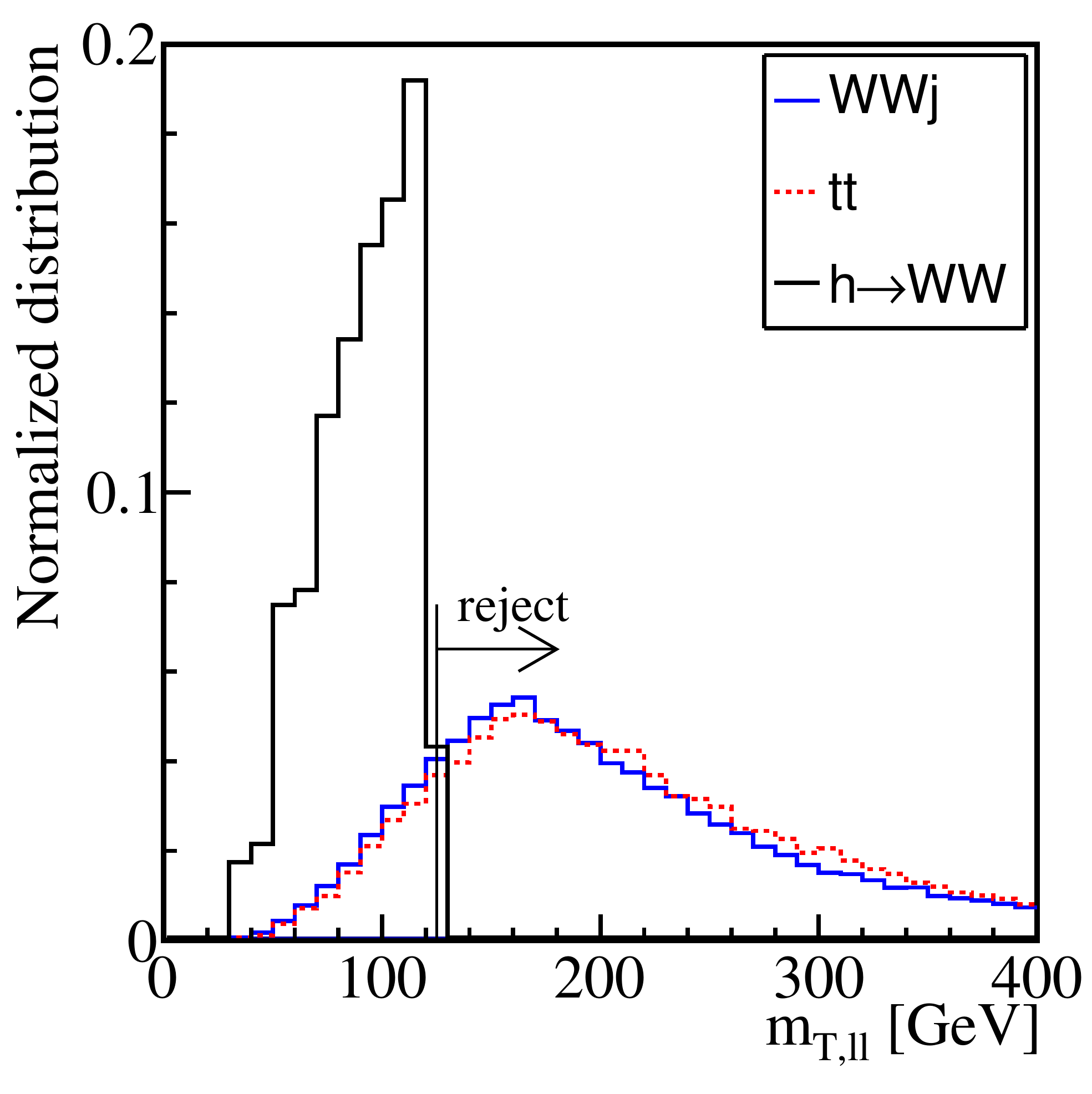}
  \includegraphics[width=0.32\textwidth]{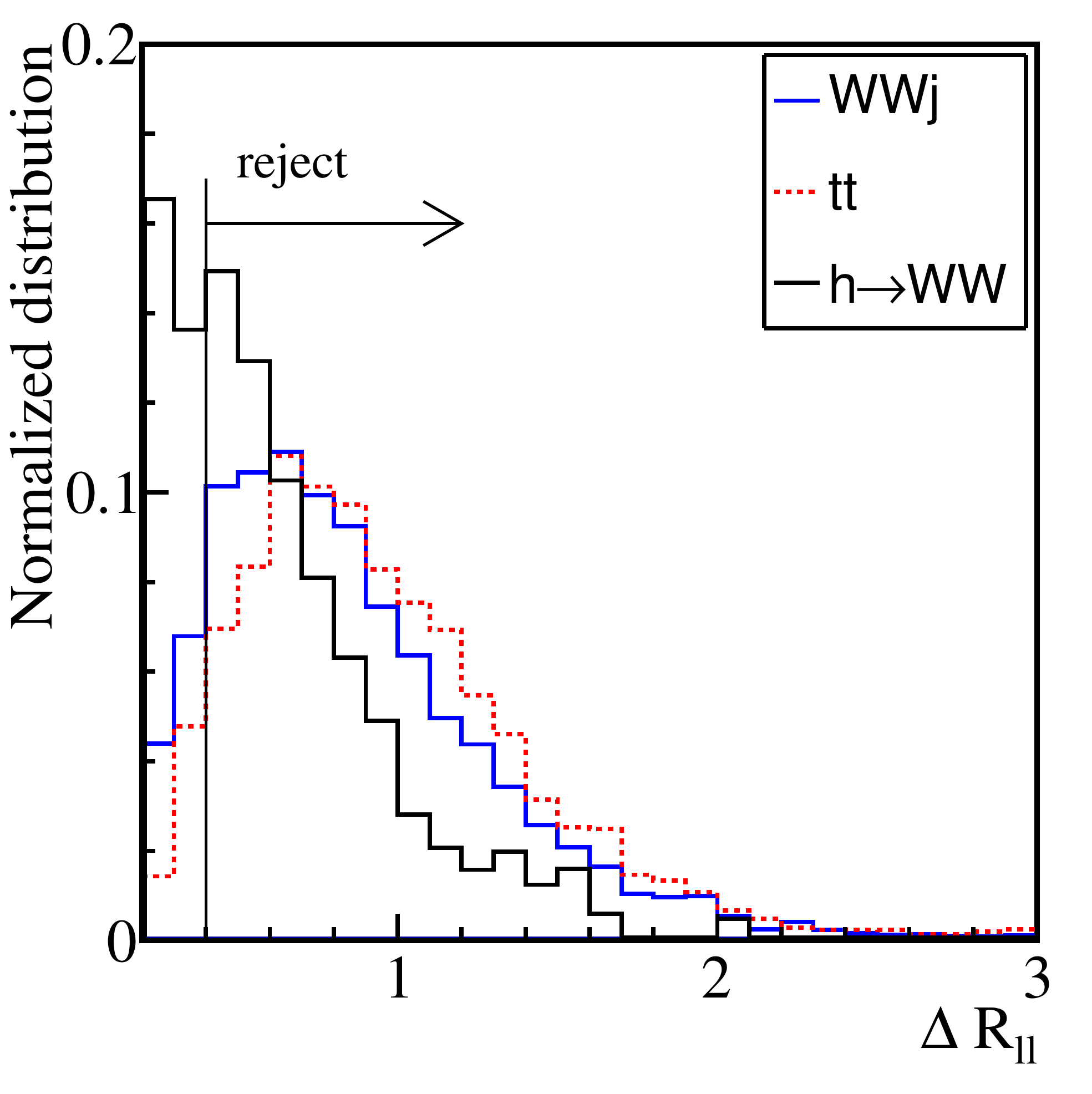}
  \includegraphics[width=0.32\textwidth]{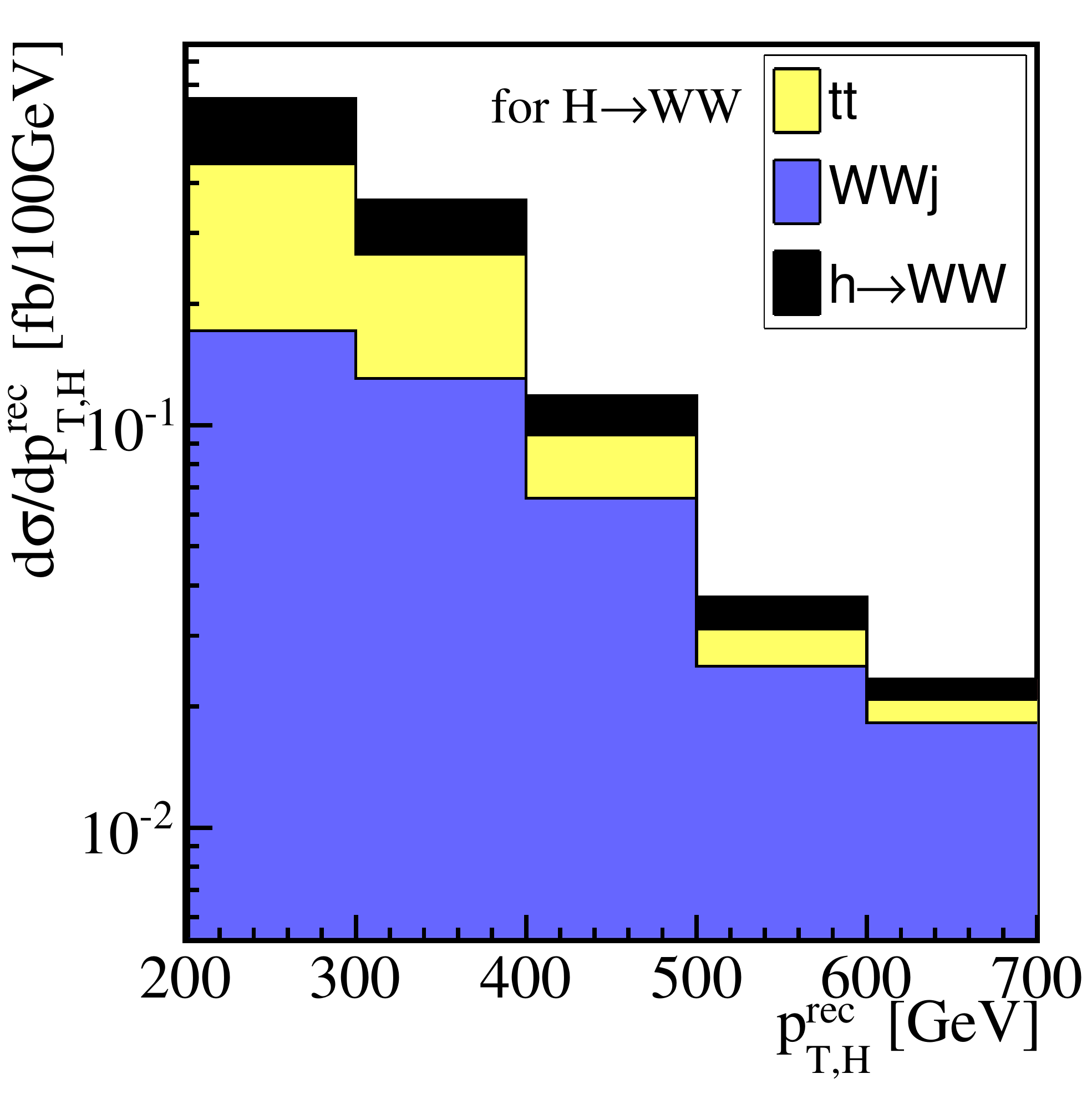}
\caption{
Distributions of the transverse mass $m_{T, \ell\ell}$ after all selection cuts up to $6'$ imposed (left panel) 
and the dilepton separation $\Delta R_{\ell\ell}$ after all selection cuts up to $7'$ imposed (central panel).
Right: stacked distribution of the `Higgs' transverse momentum $p^{\rm rec}_{T,H}$ (defined in Eq.~\eqref{eq:ptH}) 
after {\it all} selection cuts for $H \to W_\ell W_\ell^*$ optimization, with a logarithmic scale.
}
\label{fig:hww}
\end{figure}

We are now left with $H \to W_\ell W_\ell^*$ as our signal process, competing with the $WW$/$t\bar{t}$
+jets backgrounds.  Their kinematics unfortunately allow for little discriminating power: all of them contain two
leptonic $W$ bosons, with no possibility of mass reconstruction.  Luckily, the transverse mass provides
some discrimination.  As shown in~\cite{Barr:2009mx}, the transverse mass variable 
satisfying $m_{T,\ell\ell} \leq m_h$ that gives the {\it greatest} lower bound on the Higgs mass in its
decay to $W_\ell W_\ell^*$ 
\begin{equation} \label{mT} m_{T,\ell\ell}^2 = m_{\ell\ell}^2 + 2(E_{T,\ell\ell}\slashed{E}_T -
  \mathbf{p}_{T,\ell\ell}\cdot\slashed{\mathbf{p}}_T),
\end{equation}
where $E_{T,\ell\ell} = (m_{\ell\ell}^2 + p_{T,\ell\ell}^2)^{1/2}$ is the transverse energy of the dilepton
system, and $\slashed{E}_T=|\slashed{\mathbf{p}}_T|$ is the missing transverse energy.  We adopt
this definition of $m_{T,\ell\ell}$, also used by the ATLAS
Collaboration~\cite{ATLAS-CONF-2013-030}.\footnote{Setting the dilepton mass $m_{\ell\ell}$ to zero in
Eq. \eqref{mT}, despite its non-zero value being measured, gives the transverse mass used by the
CMS Collaboration~\cite{Chatrchyan:2013iaa}; this results in a less steep end point.} The end point
at $m_H$ for the transverse mass of the signal is shown in the left panel of Fig.~\ref{fig:hww},
where all the selection cuts up to step $6'$ in Table~\ref{tab:hww} have been applied. We therefore impose
\begin{equation}
  m_{T,\ell\ell} < m_H = 125~{\rm GeV}.
\end{equation}

Finally, backgrounds are further suppressed by requiring that the leptons have similar directions,
\begin{equation}
  \Delta R_{\ell \ell} < 0.4,
\end{equation}
which is typically the case for the signal due to the aforementioned spin correlations.

The efficiencies of the cuts aimed at $H \to W_\ell W_\ell^*$ are shown in Table~\ref{tab:hww}, together with the last common cut --
the $b$ veto.  We finally find $S/B \sim 0.4$,
with $S/\sqrt{B} > 6$ for 300 fb$^{-1}$.  The table also shows the event numbers left after
increased $p_T$ cuts on the reconstructed Higgs.  The resulting reconstructed Higgs $p_{T,H}^{\rm
  rec}$ distributions are shown in Fig.~\ref{fig:hww} (right panel), stacked with the signal
and background processes. As $p_{T,H}^{\rm rec}$ increases, the signal over background ratio drops faster
for the $WW$ mode selection than for the $\tau\tau$ selection.

\begin{table}[h!]
\begin{center}
\renewcommand{\arraystretch}{1.3}
  \begin{tabular}{l|cc|ccc||cc}
    Event rate [fb] & $H \to WW^*$ & $H\to \tau\tau$ &
    $W_\ell W_\ell$+jets & $Z_{\to\tau\tau}$+jets & $t_\ell\bar{t}_\ell$+jets & $S/B$ & $S/\sqrt{B}$ \cr
    \hline
    5. $n_b=0$ & 1.825 & 0.940 & 48.855 & 57.068 & 105.851 & 0.01 & 3.29\cr
    \hline
    $6^\prime$. $m_{T2}^{\ell\ell}>10$~GeV  & 1.096 & 0.002 & 25.241 & 0.028 & 53.730 & 0.01 & 2.14\cr
    $7^\prime$. $m_{T,\ell\ell}<125$~GeV & 1.095 & 0.002 & 3.809 & 0.023 & 7.235 & 0.10 & 5.70\cr
    $8^\prime$. $\Delta R_{\ell\ell}<0.4$  & 0.330 & 0.000 & 0.426 & 0.002 & 0.450 & 0.38 & 6.11 \cr
    \hline
    \ \ \ \ $p_{T,H}^{\rm rec}>300$~GeV    & 0.128 & 0.000 & 0.254 & 0.002 & 0.175 & 0.30 & 3.38 \cr
    \ \ \ \ $p_{T,H}^{\rm rec}>400$~GeV   & 0.034 & 0.000 & 0.124 & 0.000 & 0.040 & 0.21 & 1.44 \cr
    \ \ \ \ $p_{T,H}^{\rm rec}>500$~GeV  & 0.010 & 0.000 & 0.058 & 0.000 & 0.011 & 0.14 & 0.65\cr
    \ \ \ \ $p_{T,H}^{\rm rec}>600$~GeV  & 0.004 & 0.000 & 0.033 & 0.000 & 0.005 & 0.10 & 0.33 \cr
    \hline
  \end{tabular}
  \caption{Cut efficiencies for our analysis aimed at $H \to W_\ell W_\ell^*$, continued from the first part of Table~\ref{tab:htautau}.
    The values for each process are cross-sections in fb. $S/\sqrt{B}$ has been calculated for ${300~{\rm fb}^{-1}}$.}
  \label{tab:hww}
  \end{center}
\end{table}

\section{Discussion}
\label{sec:discussion}

\begin{figure}[h!]
  \includegraphics[width=0.3\textwidth]{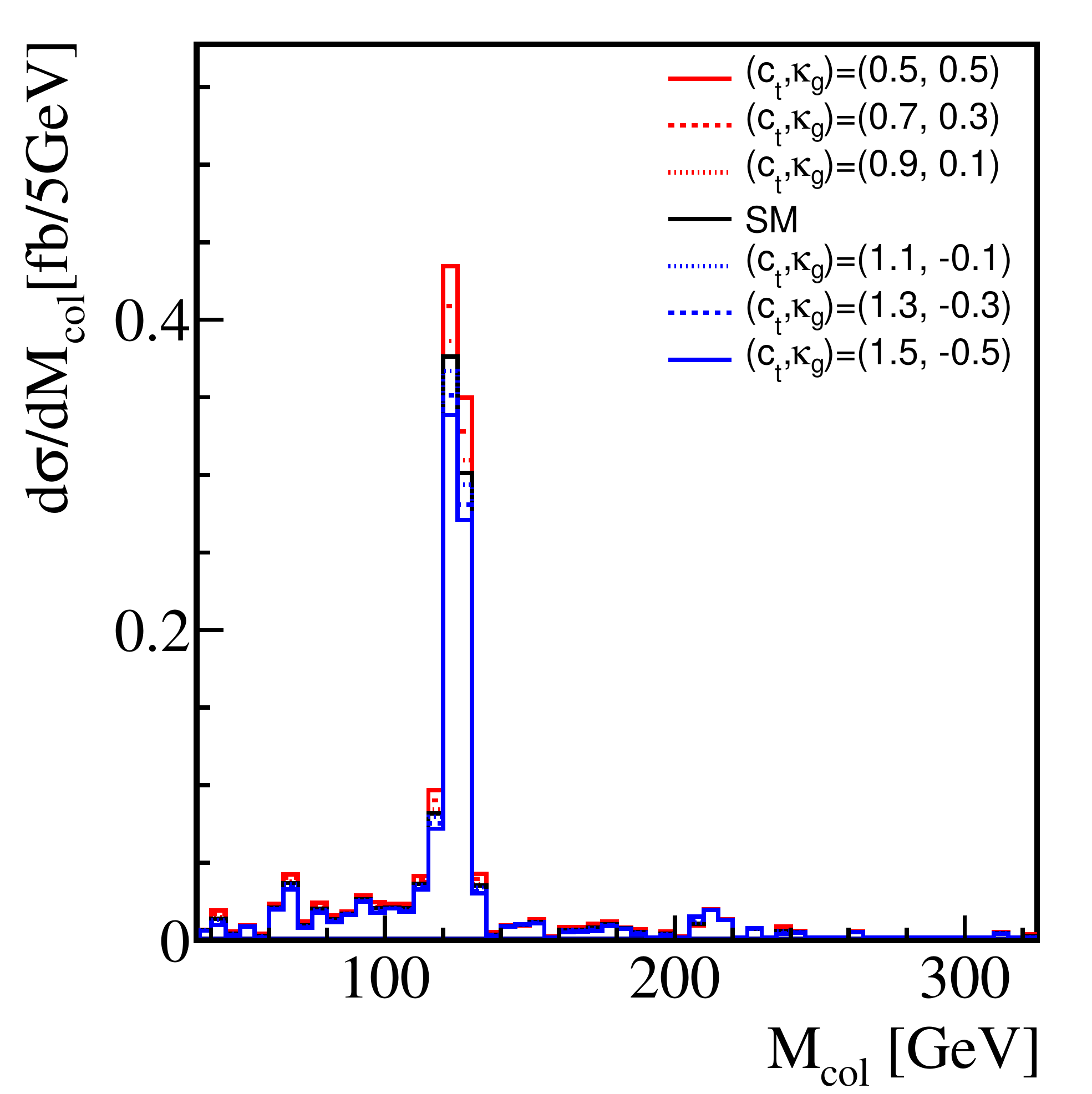}
  \includegraphics[width=0.3\textwidth]{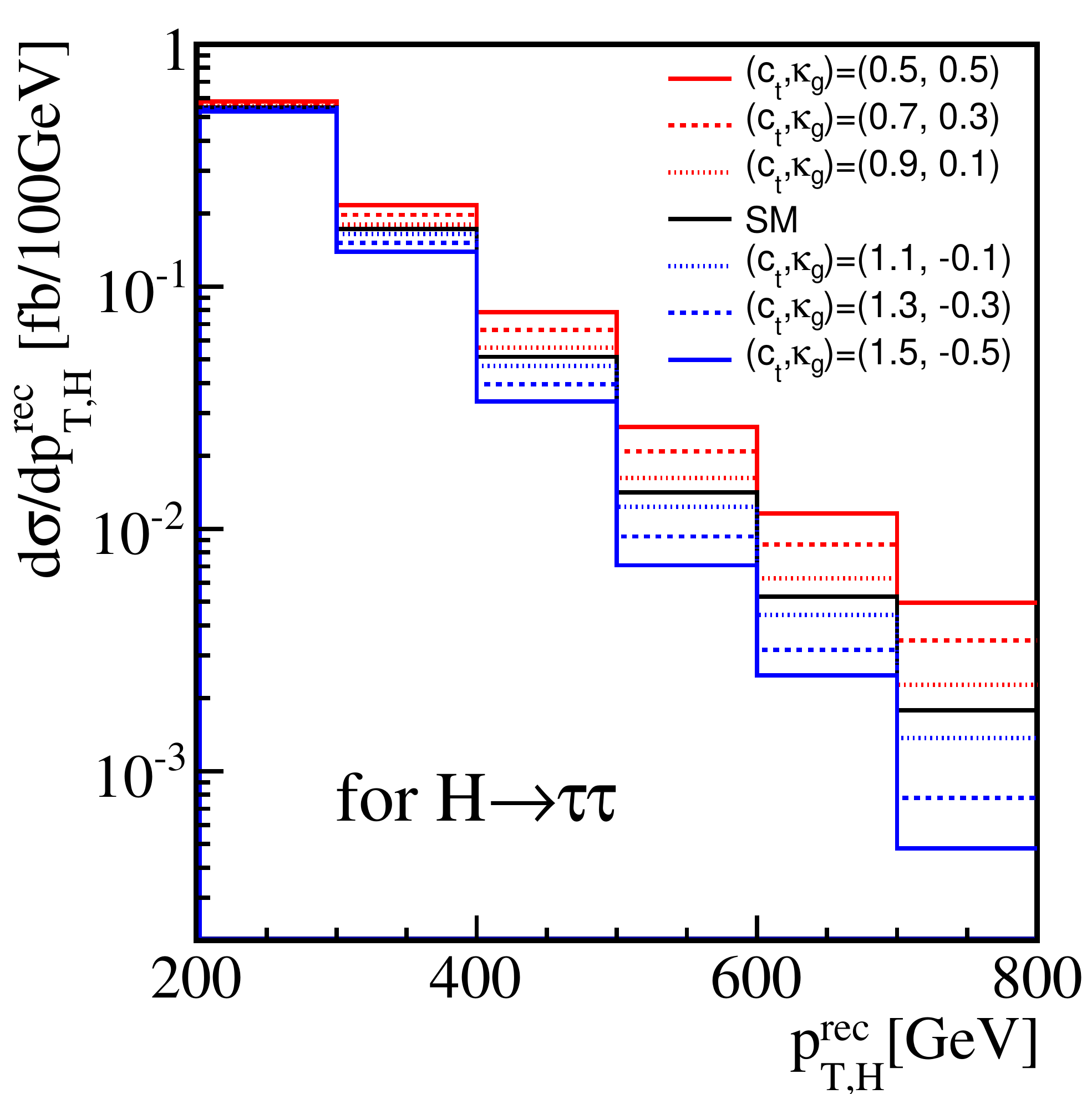}
  \includegraphics[width=0.3\textwidth]{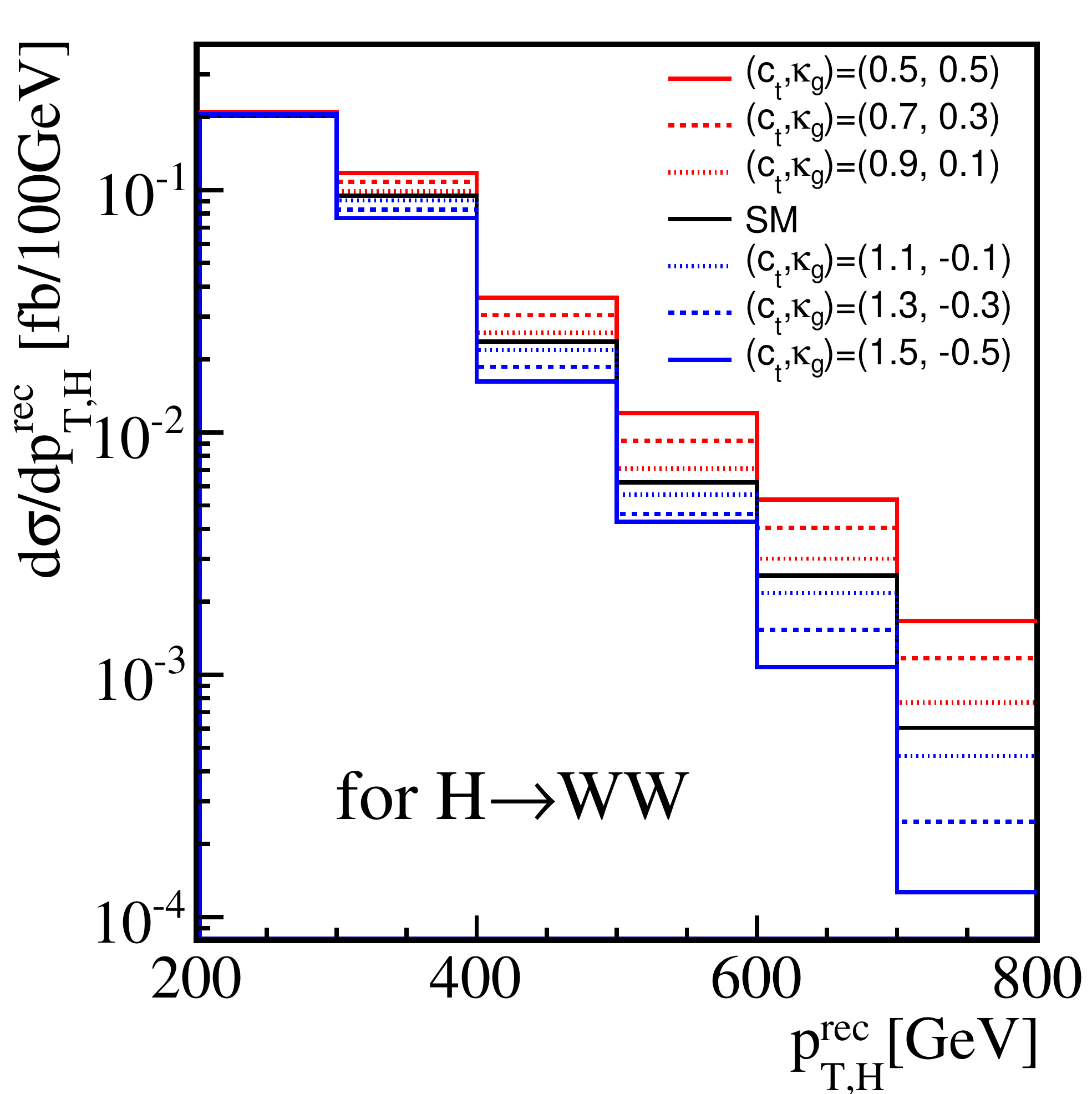}
  \caption{Signal distributions for the SM and six model points, normalized to the respective cross-sections.
  Left panel: the collinear mass $M_{\rm col}$ after cut 7.
     $p_{T,H}^{\rm rec}$ is shown for $H\to \tau\tau$
    ($H \to W_\ell W_\ell$) in the central (right) panel after all optimized selection cuts.  }
  \label{mcol_NP}
\end{figure}

In this section, we discuss how much of the difference in the $p_{T, H}$ distributions due to the modified couplings can be observed
after the realistic reconstruction of the previous section has been performed.  The left panel of
Fig.~\ref{mcol_NP} shows the signal $M_{\rm col}$ distributions for the model points after
applying the analysis described in Sec.~\ref{TauTau} up to cut 7.  We see the peak in the observable for all
points.  The central and right panel show the signal $p_{T,H}^{\rm rec}$ distributions
after the reconstruction described in Sec.~\ref{sec:SM} for $H\to \tau\tau$ and $H \to W_\ell
W_\ell$ optimizations, respectively.  As we expect, the difference in shape expected from the
parton level result of Fig.~\ref{fig:NPmodels} manifests itself also in the reconstructed $p_{T,H}^{\rm
  rec}$ distributions. A detailed breakdown after successive selection cuts
is shown in Table~\ref{tab:tautauNPratio} for the $H\to \tau\tau$ optimization and in
Table~\ref{tab:wwNPratio} for the $H\to W_\ell W_\ell$ optimization, quoting cross-sections relative 
to the corresponding SM value.  Compared with the parton level numbers in
Table~\ref{NPpartonlevel}, the $p_{T,H}^{\rm rec}$ dependence is more enhanced at the reconstructed
level. This is because most of the selection cuts are more efficient for the boosted Higgs
event topology.  \bigskip

\begin{table}[h!]
\begin{center}
\renewcommand{\arraystretch}{1.2}
  \begin{tabular}{l|ccccc|c|ccccc}
    Model point ($\kappa_g$) & 0.5 & 0.4 &  0.3 &  0.2 &  0.1 & 0 ({SM}) &  -0.1 &  -0.2 &  -0.3 &  -0.4 & -0.5 \cr 
    \hline
    3. $p_{T,H}^{\rm rec} > 200$~GeV   & 1.109 & 1.084 & 1.061 & 1.039 & 1.019 & 1.000 & 0.983 & 0.968 & 0.954 & 0.942 & 0.932\cr
    4. $n_j^{\rm fat}=1$  & 1.143 & 1.110 & 1.079 & 1.050 & 1.024 & 1.000 & 0.978  & 0.959 & 0.941 & 0.926 & 0.913\cr
    5. $n_b=0$ & 1.143 & 1.110 & 1.079 & 1.050 & 1.024  & 1.000 & 0.978& 0.959 & 0.941 & 0.926 & 0.913\cr
    \hline
    6. $\slashed{\mathbf{p}}_T$ inside two $\ell$s & 1.156 & 1.120 & 1.086 & 1.055 & 1.026 & 1.000 & 0.976 & 0.954 & 0.935 & 0.918 & 0.903\cr
    7. $m_{\ell\ell} < 70$~GeV  & 1.157 & 1.121 & 1.087 & 1.056 & 1.027 & 1.000 & 0.976 & 0.954 & 0.934  & 0.917 & 0.902\cr
    8. $|M_{\rm col}  - m_H| < 10$~GeV  & 1.163 & 1.125 & 1.091 & 1.058 & 1.028 & 1.000& 0.974 & 0.951  & 0.930 & 0.912 & 0.896\cr
    \hline
    \ \ \ \ $p_{T,H}^{\rm rec}>300$~GeV & 1.392 & 1.303 & 1.219 & 1.140 & 1.067 & 1.000 & 0.938 & 0.882 & 0.831 & 0.785 & 0.745\cr
    \ \ \ \ $p_{T,H}^{\rm rec}>400$~GeV & 1.711 & 1.544 & 1.389 & 1.247 & 1.117 & 1.000 & 0.895 & 0.802 & 0.722 & 0.653 & 0.597\cr
    \ \ \ \ $p_{T,H}^{\rm rec}>500$~GeV & 2.131 & 1.857 & 1.607 & 1.381 & 1.179 & 1.000 & 0.845 & 0.715 & 0.608 & 0.525 & 0.465\cr
    \ \ \ \ $p_{T,H}^{\rm rec}>600$~GeV & 2.602 & 2.201 & 1.840 & 1.520 & 1.240 & 1.000 & 0.801 & 0.642 & 0.523 & 0.445 & 0.407\cr
    \hline
  \end{tabular}
  \caption{The relative cross-section $\sigma/\sigma_{\rm SM}$ for several new physics model points after successive selection cuts for $\tau\tau$ optimization.}
  \label{tab:tautauNPratio}
  \end{center}
\end{table}

\begin{table}[h!]
\begin{center}
\renewcommand{\arraystretch}{1.2}
  \begin{tabular}{l|ccccc|c|ccccc}
    Model point ($\kappa_g$) & 0.5 & 0.4 &  0.3 &  0.2 &  0.1 & 0 ({SM}) &  -0.1 &  -0.2 &  -0.3 &  -0.4 & -0.5 \cr 
    \hline
    5. $n_b=0$  & 1.143 & 1.110 & 1.079 & 1.050 & 1.024 & 1.000 & 0.978 & 0.959 & 0.941 & 0.926 & 0.913\cr
    \hline
    $6^\prime$. $m_{T2}^{\ell\ell}>10$~GeV   & 1.117 & 1.089 & 1.064 & 1.040 & 1.019 & 1.000 & 0.983 & 0.968 & 0.955 & 0.944 & 0.936\cr
    $7^\prime$. $m_{T,\ell\ell}<125$~GeV & 1.117 & 1.089 & 1.064 & 1.040 & 1.019 & 1.000 & 0.983 & 0.968 & 0.955 & 0.944 & 0.936\cr
    $8^\prime$. $\Delta R_{\ell\ell}<0.4$ & 1.164 & 1.125 & 1.088 & 1.056 & 1.026 & 1.000 & 0.977 & 0.958 & 0.942 & 0.929 & 0.920\cr
    \hline
    \ \ \ \ $p_{T,H}^{\rm rec}>300$~GeV  & 1.360 & 1.278 & 1.201 & 1.129 & 1.062 & 1.000 & 0.943 & 0.892 & 0.845 & 0.803 & 0.767\cr
    \ \ \ \ $p_{T,H}^{\rm rec}>400$~GeV & 1.684 & 1.520 & 1.370 & 1.234 & 1.110 & 1.000 & 0.903 & 0.819 & 0.749 & 0.692 & 0.648\cr
    \ \ \ \ $p_{T,H}^{\rm rec}>500$~GeV & 2.093 & 1.822 & 1.577 & 1.358 & 1.166 & 1.000 & 0.860 & 0.747 & 0.659 & 0.598 & 0.564\cr
    \ \ \ \ $p_{T,H}^{\rm rec}>600$~GeV & 2.377 & 2.043 & 1.738 & 1.463 & 1.217 & 1.000 & 0.812 & 0.654 & 0.525 & 0.425 & 0.354\cr
    \hline
  \end{tabular}
  \caption{Relative size $\sigma/\sigma_{\rm SM}$ for several new physics model points after successive selection cuts 
    for $WW$ optimization.}
  \label{tab:wwNPratio}
  \end{center}
\end{table}

We will now estimate how much integrated luminosity is needed to find a certain 
significance for the signal.
We perform a binned likelihood analysis of signal and background using the $\mathrm{CL}_s$ method, as described in \cite{Junk:1999kv}. We include systematic errors on the cross-section normalization assuming a Gaussian probability distribution.

\begin{figure}[h!]
  \includegraphics[width=0.32\textwidth]{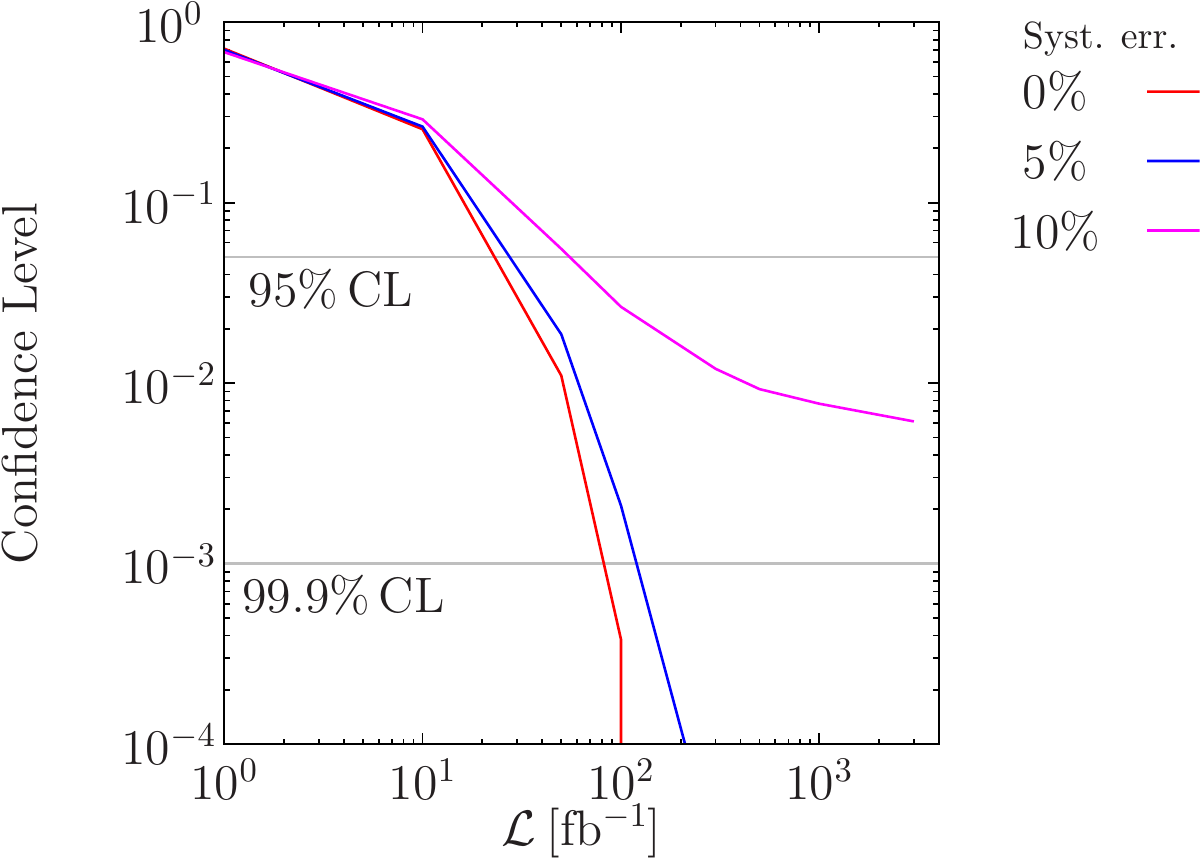}
  \includegraphics[width=0.32\textwidth]{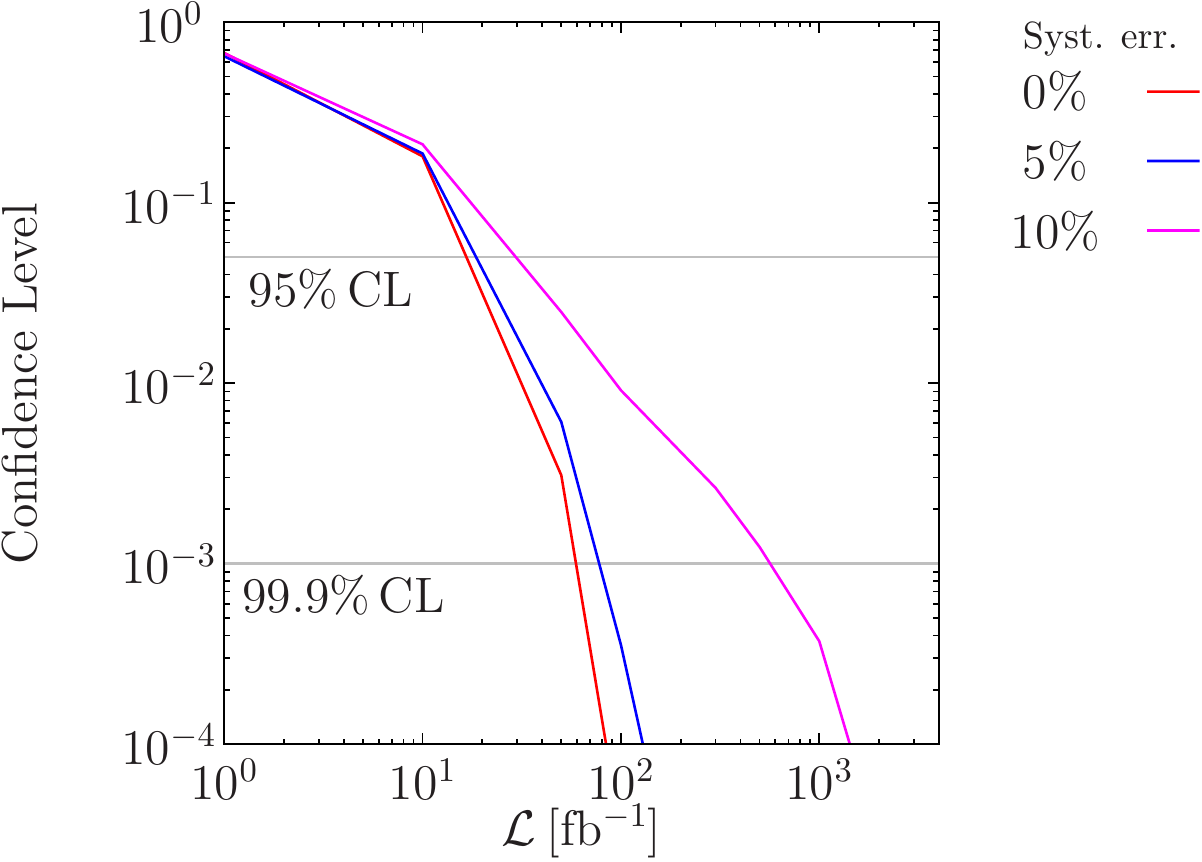}
  \includegraphics[width=0.32\textwidth]{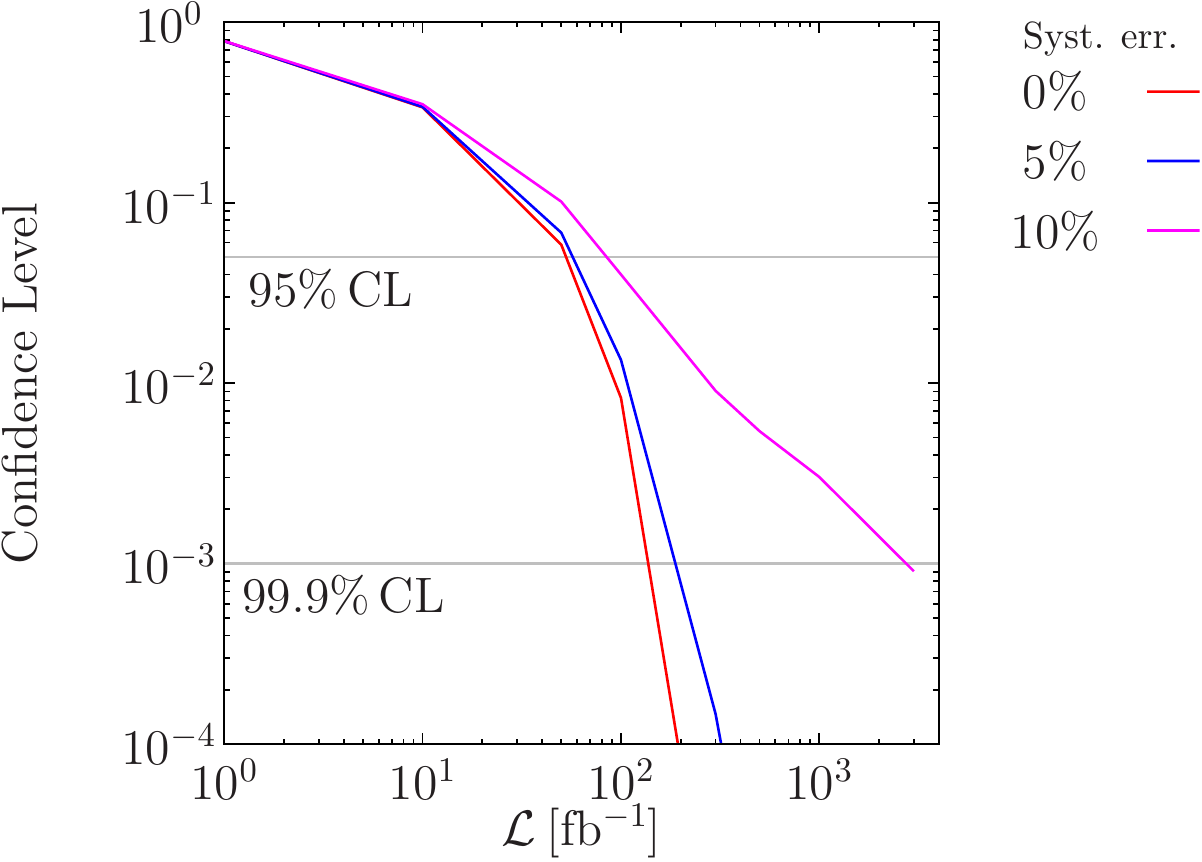}
  \caption{CL$_s$ vs. the integrated luminosity for the model points  $\kappa_g= 0$ (SM, left) and  $\kappa_g=0.5$ (central) against a background-only hypothesis using the $\tau\tau$ mode.
  Right panel: CL$_s$ plot for the model point of $\kappa_g=0.5$ against a background-only hypothesis using the $WW$ mode.  }
  \label{CLs_BG}
\end{figure}

Fig.~\ref{CLs_BG} shows the expected $p$-values as a function of 
the integrated luminosity ${\cal L}$ in the SM (left panel), the model point of $\kappa_g=0.5$ (central panel) using the $H \to \tau\tau$ analysis and $\kappa_g=0.5$ using the $H \to W_\ell W_\ell$ analysis (right panel).
The analysis is based on the expected signal-plus-background against a background-only hypothesis.
In the analysis, three different systematic errors on the cross-section normalization of 0, 5, and 10\% are assumed. While achieving theoretical uncertainties of less than $10~\%$ is challenging, in the separation of signal and background we rely predominantly on the lepton momenta which can be measured very precisely.
As one can see from the left panel in Fig.~\ref{CLs_BG}, with ${\cal L}=20 \sim 60$ fb$^{-1}$, we are able to 
see the SM signal at 95\% confidence level depending on the assumed systematic uncertainty.

For $\kappa_g>0$, the signal is enhanced and the required integrated luminosity decreases:
it would be ${\cal L}=15 \sim 30$ fb$^{-1}$ for $\kappa_g=0.5$ to observe the signal at 95\% CL, as shown in the central panel.

The right panel of Fig.~\ref{CLs_BG} shows the $p$-values for $\kappa_g=0.5$ using the $H \to W_\ell W_\ell$ mode.
The sensitivity compared to the $\tau\tau$ mode is slightly reduced. However, it is still possible to exploit the $W_\ell W_\ell$ final state to observe a boosted Higgs boson.
 
\begin{figure}[h!]
  \includegraphics[width=0.33\textwidth]{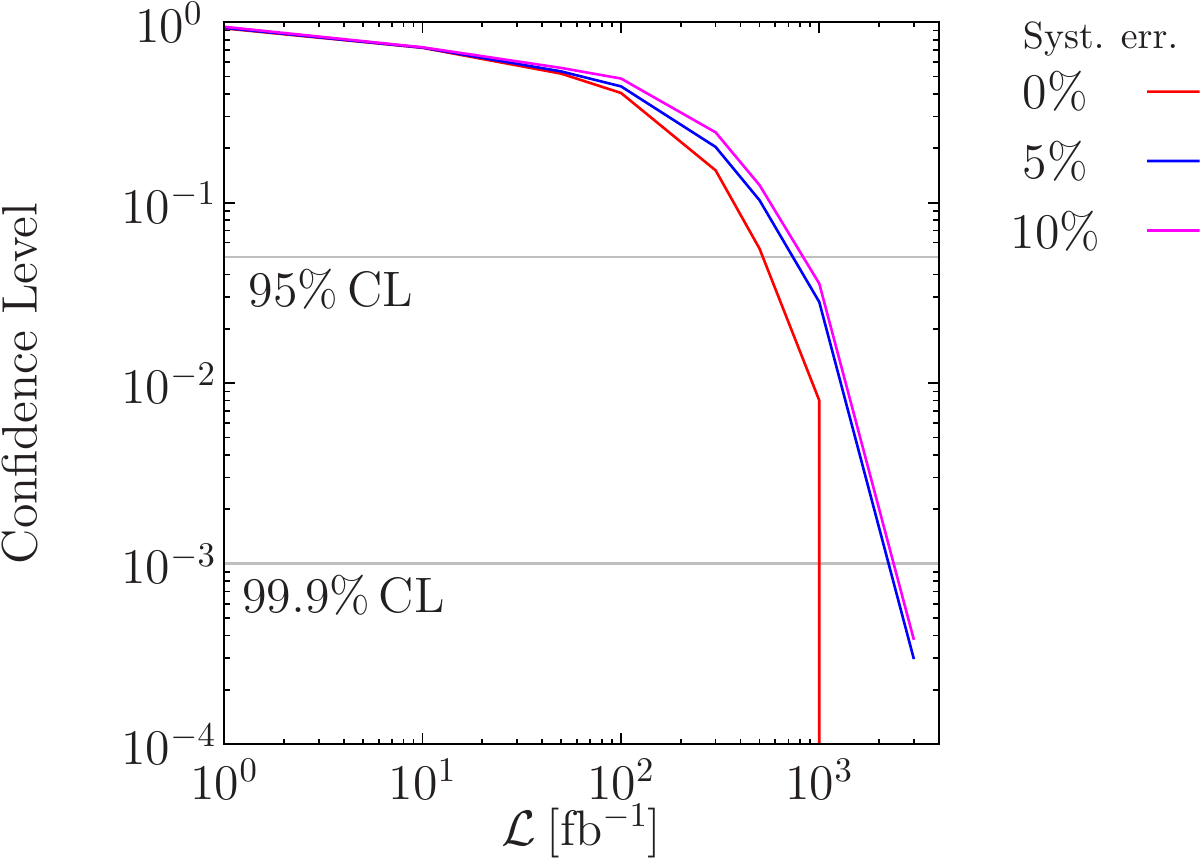}
  \includegraphics[width=0.33\textwidth]{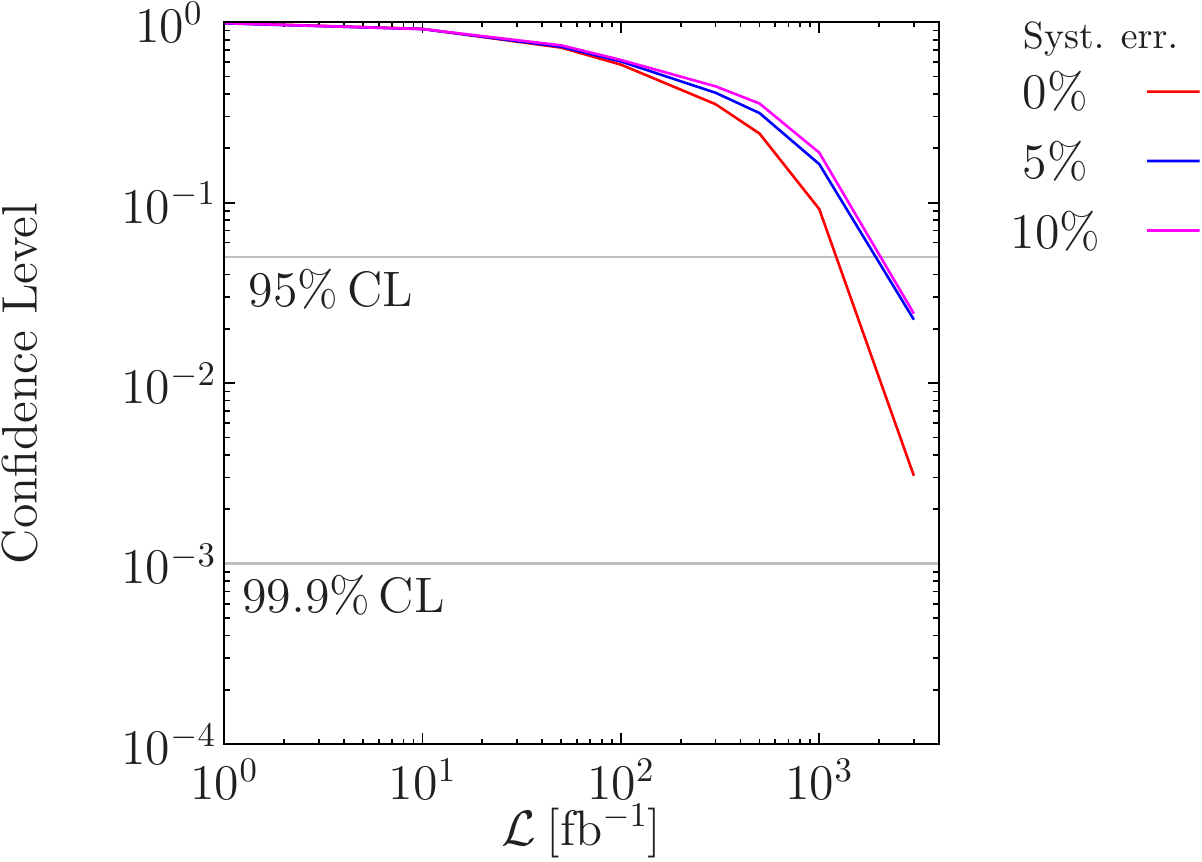}
  \includegraphics[width=0.27\textwidth]{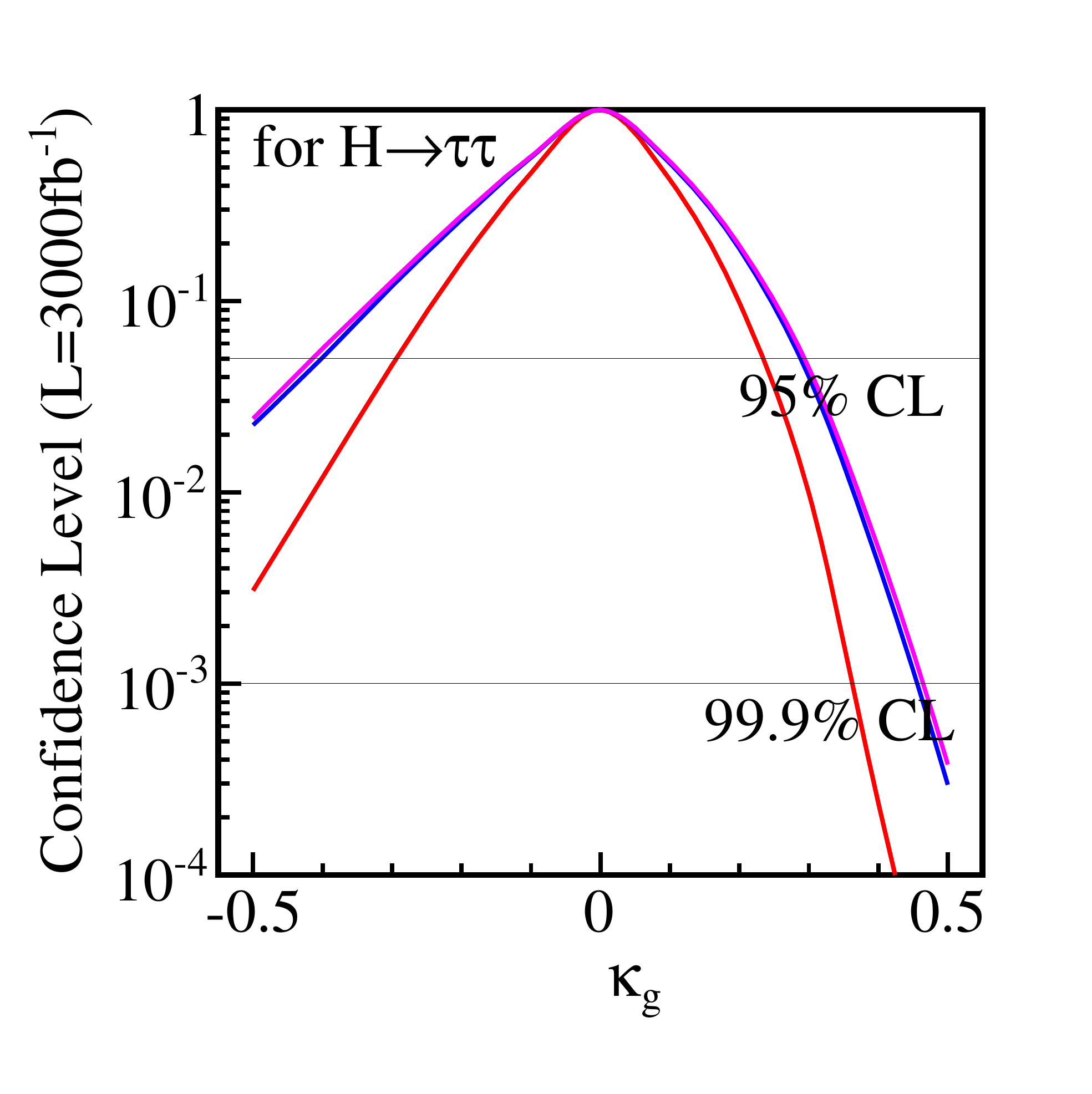}
  \caption{CL$_s$ vs. the integrated luminosity using $\tau\tau$ mode for the model point of $\kappa_g=0.5$ (left) and $\kappa_g=-0.5$ (central).
  Right: CL$_s$ as a function of $\kappa_g$ for an integrated luminosity of 3000 fb$^{-1}$.
  }
  \label{CLs}
\end{figure}

We also perform a binned likelihood analysis
to estimate how well we can distinguish these model points from the SM given the presence of backgrounds.
The left panel of Fig.~\ref{CLs} shows the expected $p$-values 
to observe the signal and background against the SM and background hypothesis
as a function of the integrated luminosity ${\cal L}$ for the model point of $\kappa_g=0.5$ using the $H \to \tau\tau$ analysis. 
Again, systematic errors of 0, 5, and 10\% are assumed.  
We find that we are able to
distinguish the model point $\kappa_g=0.5$ from the SM with ${\cal L}=1000$ fb$^{-1}$ 
even assuming 10\% systematic uncertainty.

It is more difficult to prove a deviation from the SM 
for model points with $\kappa_g < 0$, compared to $\kappa_g>0$ with the same $|\kappa_g|$ value,
since this gives a deficit rather than a surplus of signal events.
The central panel of Fig.~\ref{CLs} shows the $p$-values for $\kappa_g=-0.5$ using the $H \to \tau\tau$ analysis.
As expected we have less sensitivity, and even smaller values of $|\kappa_g|$ require larger integrated luminosities.

The right panel of Fig.~\ref{CLs} shows the $p$-values as a function of $\kappa_g$ 
using the $H \to \tau\tau$ for an integrated luminosity of 3000 fb$^{-1}$.
If we assume 0\% systematic uncertainty we can exclude $\kappa_g < -0.29$ and $\kappa_g > 0.24$ 
for ${\cal L}=3000$ fb$^{-1}$ at 95\% CL.
For the same integrated luminosity, assuming 10\% systematic uncertainty, we can still exclude $\kappa_g < -0.4$ and $\kappa_g > 0.3$ 
at 95\% CL.

We have not combined the $\tau\tau$ and $W_\ell W_\ell$ analyses although it could improve our sensitivity by some amount.
Combining both channels is a complex task since the systematic uncertainties of both channels have to be evaluated by the experimental collaborations. Furthermore, it is not easy to avoid double-counting of events when combining both decay modes, as the final state reconstructions discussed in Sec.~\ref{sec:SM} are not able to strictly separate them (see Table \ref{tab:htautau}).

\section{Conclusions} \label{conc}

The dominant production mode of the Higgs boson at the LHC -- gluon fusion -- is an important probe of new physics. Even though the inclusive rate has been measured to be in agreement with the SM, the study of a Higgs boosted by recoil against a hard jet constitutes an interesting, albeit challenging, measurement.
It is motivated in the context of supersymmetry and composite Higgs models, and indeed generically in natural new physics: the Higgs coupling to a top-quark loop is both central to the question of natural electroweak symmetry breaking, and the chief source of gluon fusion.
Due to the low energy theorem however, the details of this loop-induced process are entirely obscured unless one can access the boosted Higgs regime.

We have shown boosted Higgs signal isolation in the dilepton channel via $H \to \tau\tau$ and $H \to WW$.
The boost enhances the efficiency of the collinear approximation for mass reconstruction in the $H \to \tau\tau$ mode, giving a peak at $m_H$
visible above the dominant $Z$+jets background. $Z$+jets provides its own peak for this reconstructed mass distribution; using the sidebands around the $m_H$ peak we expect a
relatively precise background estimate. In the end we achieve $S/B\sim 0.4$.
For $H \to WW$ mode, we can also achieve $S/B\sim 0.4$ but with fewer events.
This is nevertheless a helpful addition to the statistical significance. We expect a 12\% error for the cross-section measurement for 
$p_T>200$~GeV, 22\% for $p_T>300$~GeV, and 41\% for $p_T>400$~GeV with an integrated luminosity of 300~fb$^{-1}$.

A direct measurement of the top Yukawa coupling in the $t\bar{t}H$ channel is also instrumental for breaking the degeneracy concerning
the coupling of the Higgs to gluons and to the top quark, and the $H+$jets mode provides a complementary determination.
We have shown that we can distinguish several new physics models in an
effective field theory approach using the reconstructed Higgs $p_T$ distribution. 
With an integrated luminosity of 3000 fb$^{-1}$ at the 14 TeV LHC, we can exclude 
$\kappa_g < -0.4$ and $\kappa_g > 0.3$ along the line $c_t + \kappa_g=1$
at 95\% confidence level assuming the systematic uncertainty of 10\%.

\section{Acknowledgments}
We thank Christophe Grojean and Ennio Salvioni for many helpful discussions. CW gratefully
acknowledges additional funding from the IPPP beyond his studentship, and from the French ANR,
Project DMAstroLHC, ANR-12-BS05-0006. MS acknowledges the hospitality at the CERN TH division and
the funding of his work by the Joachim Herz Stiftung.  The work of AW was supported in part by the
German Science Foundation (DFG) under the Collaborative Research Center (SFB) 676.
This work was supported in part by the STFC.\\

\clearpage \newpage

\appendix
\section{$M_{\rm col}$ distributions and data-driven background estimates}
\label{appendix}

\begin{figure}[h!]
\includegraphics[width=0.3\textwidth]{mcol_log}
\includegraphics[width=0.3\textwidth]{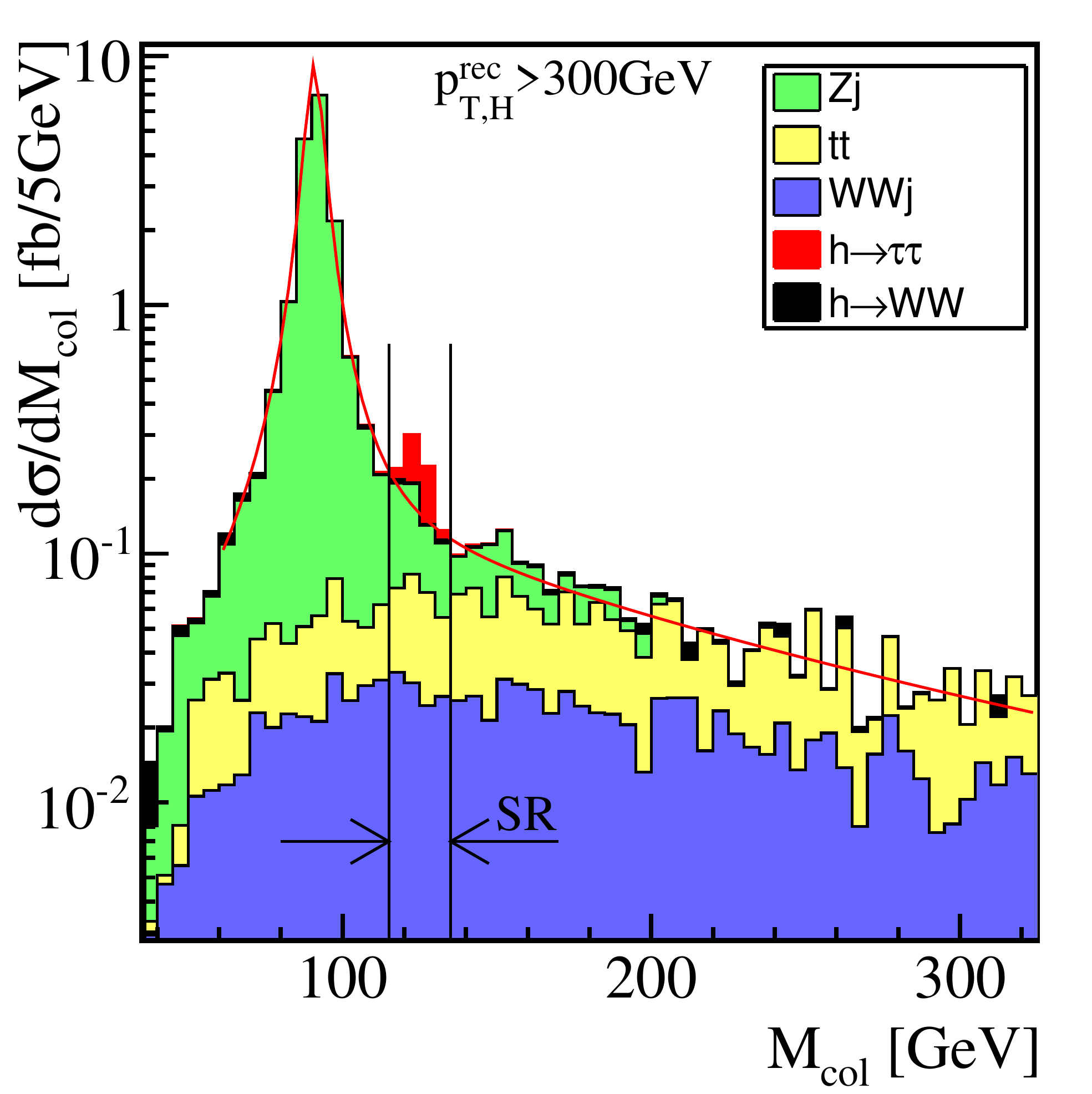}
\includegraphics[width=0.3\textwidth]{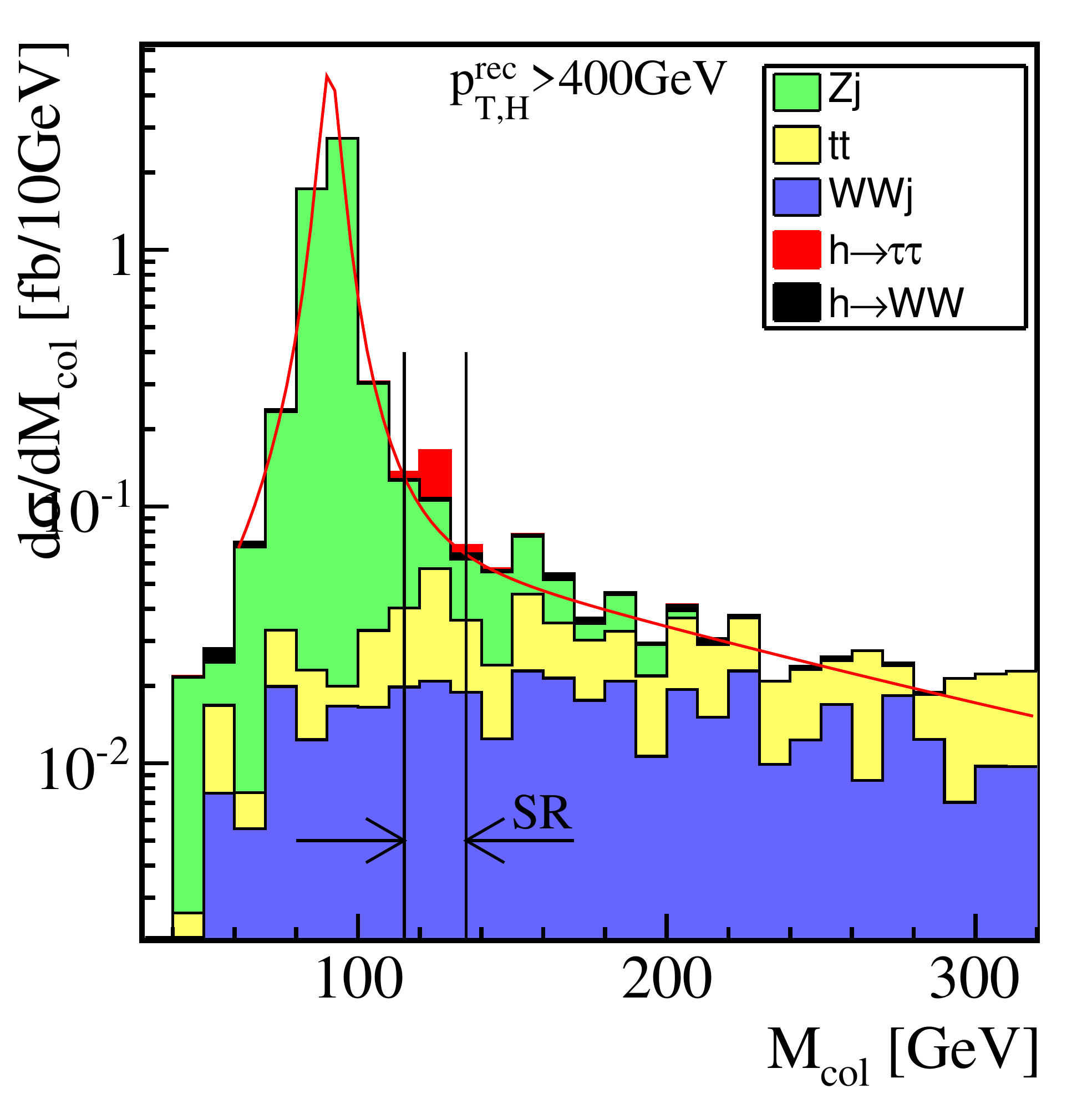}

\includegraphics[width=0.3\textwidth]{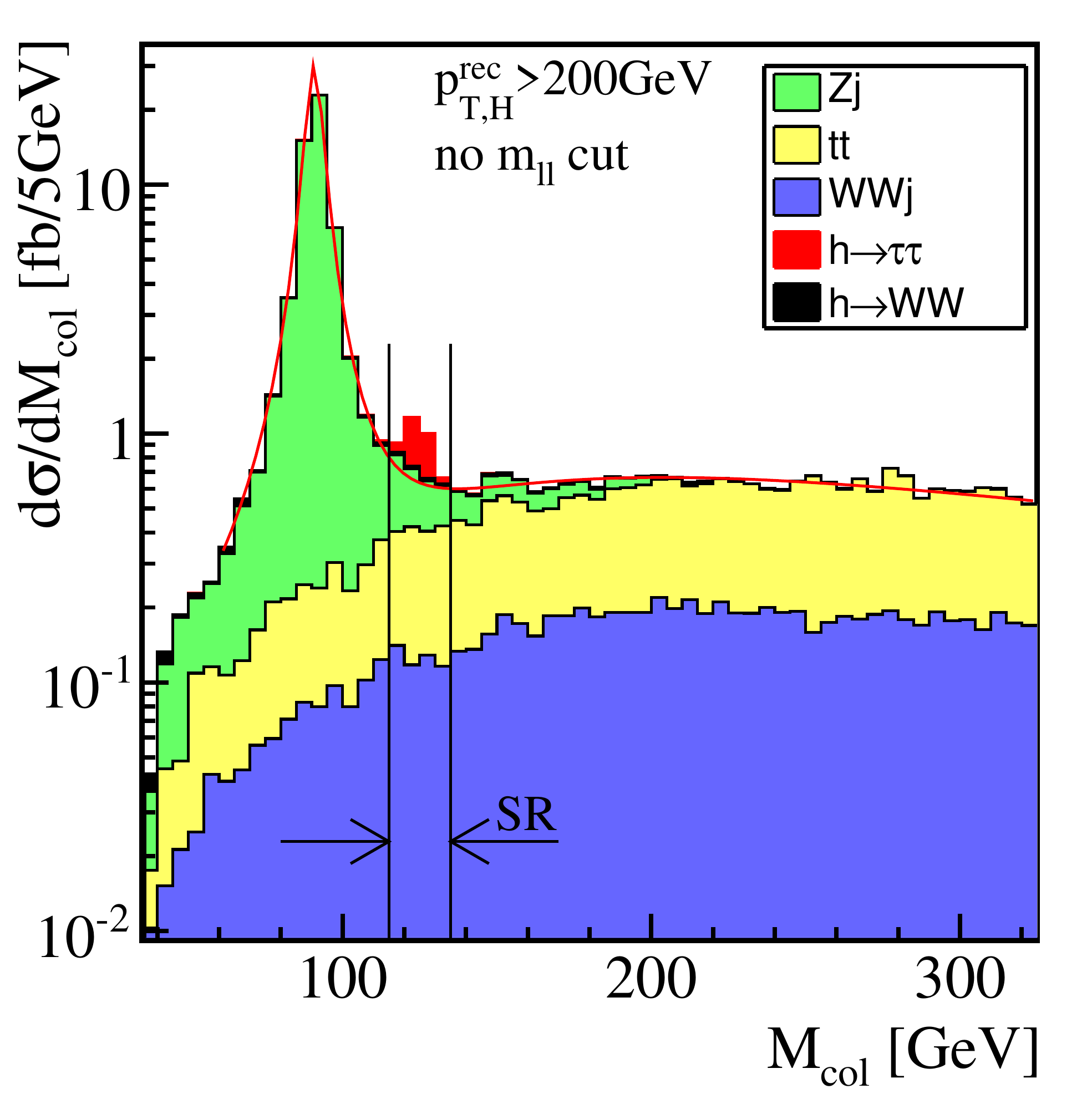}
\includegraphics[width=0.3\textwidth]{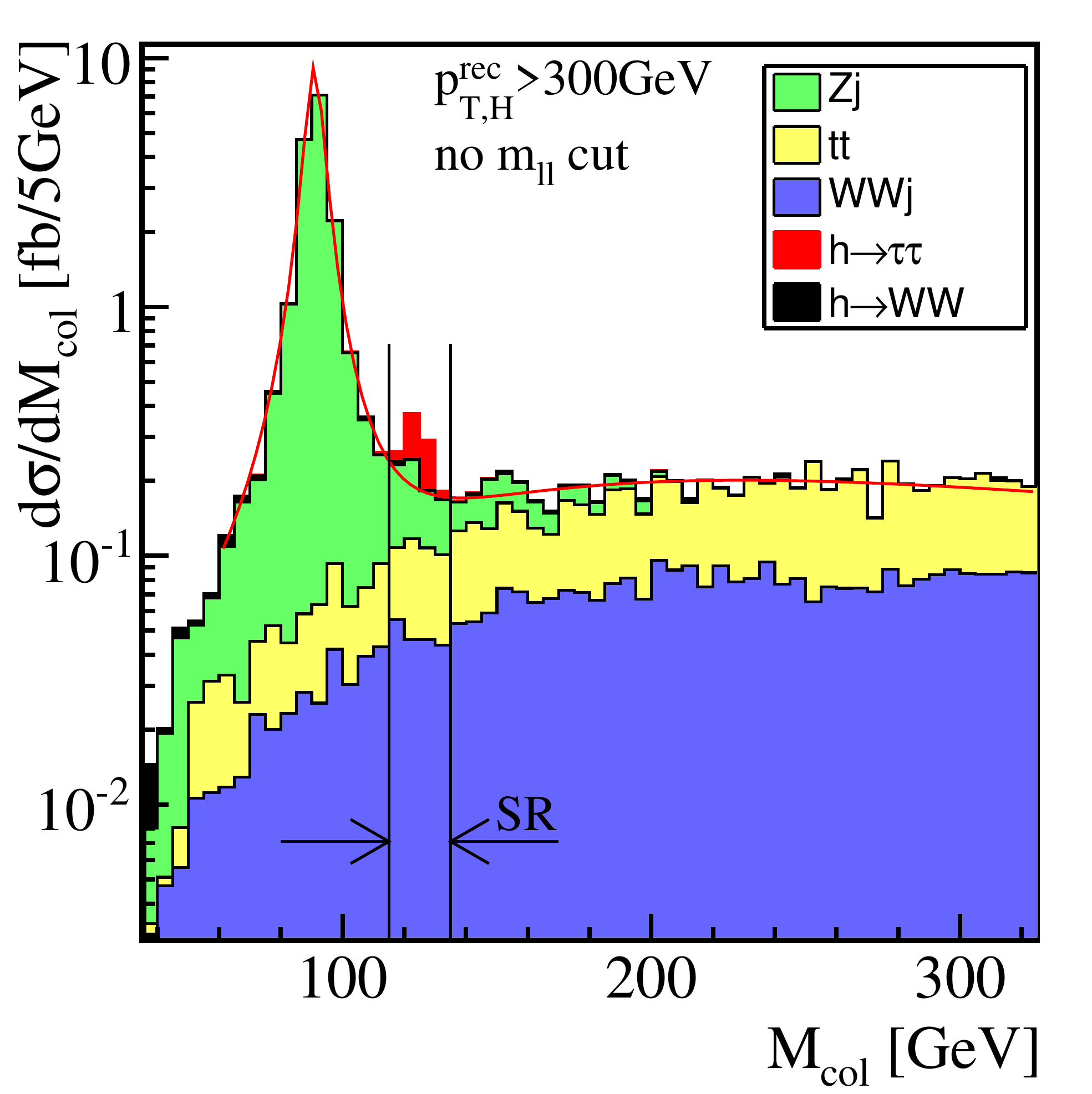}
\includegraphics[width=0.3\textwidth]{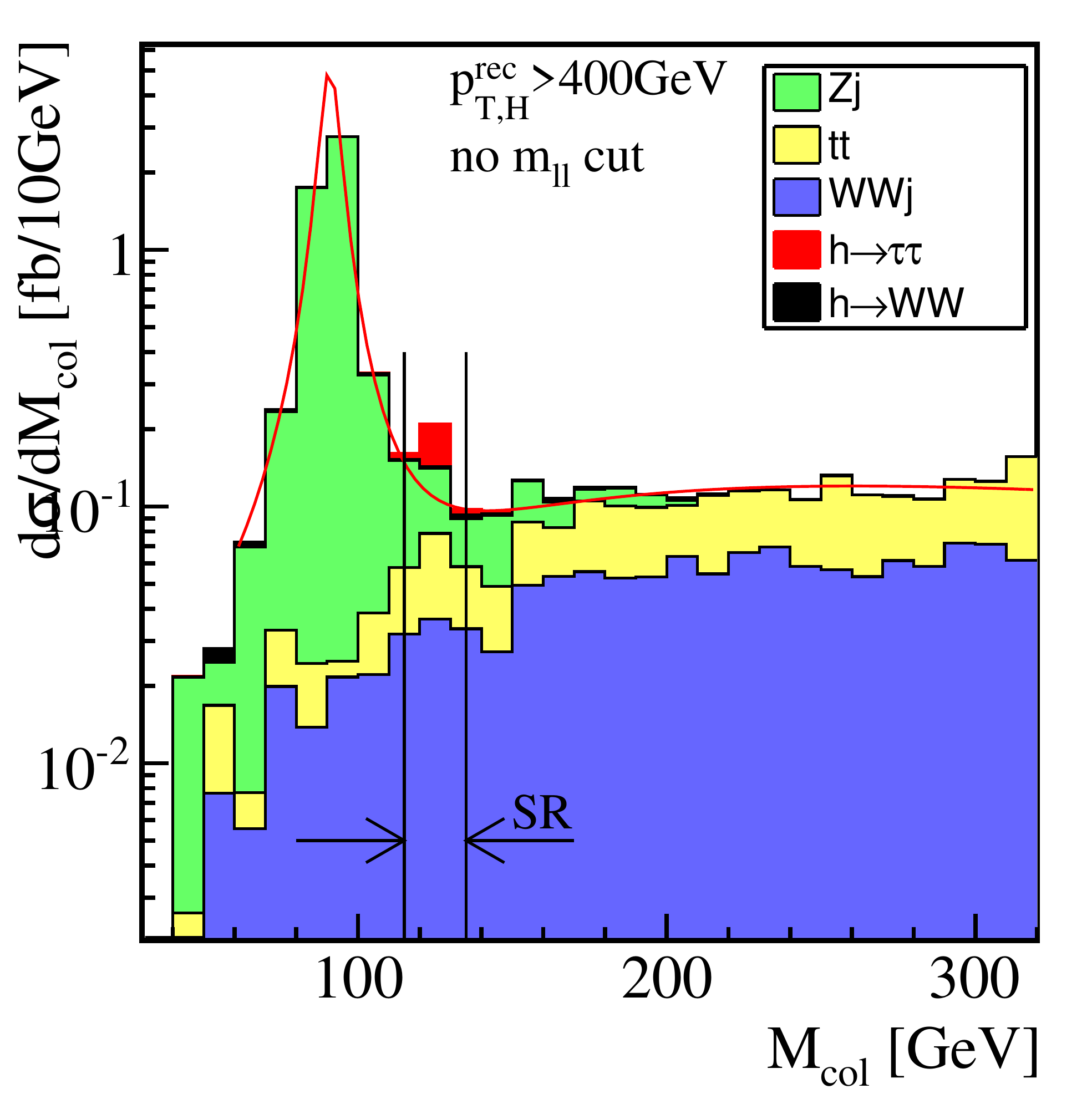}
\caption{
The collinear mass $M_{\rm col}$ distributions after (upper) and before (lower) cut 7, stacking the different processes.
Histograms are normalized to the respective cross-sections.
The $p_T$ cut on the reconstructed Higgs are $p_{T,H}^{\rm rec}>200$~GeV, $p_{T,H}^{\rm rec}>300$~GeV, and $p_{T,H}^{\rm rec}>400$~GeV from left to right.
}
\label{mcol_app}
\end{figure}

We collect distributions of the collinear mass $M_{\rm col}$  for several minimal values of the reconstructed Higgs $p_T$ and discuss how a data-driven background estimate could be performed. In Fig.~\ref{mcol_app}, we show distributions of $M_{\rm col}$ for $p_{T,H}^{\rm rec}>200$~GeV, $300$~GeV, and $400$~GeV. The upper three plots include the selection cuts up to step 7 in Table~\ref{tab:htautau}, while the lower three plots are up to cut 6 (i.e. without the $m_{\ell\ell}$ cut).
The red lines show the fitting curves for the background distributions. We take the fitting function as the sum of a Breit-Wigner function and a log-normal function.
As one can see, the $Z$-peak and the tail distributions are well fitted for a wide $p_{T,H}^{\rm rec}$ range. 
This means we can estimate the contributions of the background processes using side bands, which reduces the sensitivity to theoretical uncertainties.

Moreover, the lower plots without the $m_{\ell\ell}$ cut have larger $t\bar{t}$ and $WW$
contributions but are still well fitted with the same fitting function.
Thus, we can extract the normalizations of $t\bar{t}$ and $WW$ contributions and control part of the Monte Carlo uncertainties using data.
We therefore only consider the statistical uncertainty of the total background contributions in the signal region in the main text.

\bibliographystyle{h-physrev3}
\bibliography{TheBib}

\begin{thebibliography}{100}

\bibitem{Aad:2012tfa}
ATLAS Collaboration, G.~Aad {\em et~al.},
\newblock Phys.Lett. {\bf B716}, 1 (2012), 1207.7214.

\bibitem{Chatrchyan:2012ufa}
CMS Collaboration, S.~Chatrchyan {\em et~al.},
\newblock Phys.Lett. {\bf B716}, 30 (2012), 1207.7235.

\bibitem{Aad:2013xqa}
ATLAS Collaboration, G.~Aad {\em et~al.},
\newblock Phys.Lett. {\bf B726}, 120 (2013), 1307.1432.

\bibitem{Bolognesi:2012mm}
S.~Bolognesi {\em et~al.},
\newblock Phys.Rev. {\bf D86}, 095031 (2012), 1208.4018.

\bibitem{ATLAS-CONF-2013-029}
ATLAS Collaboration,
\newblock CERN Report No. ATLAS-CONF-2013-029, 2013 (unpublished).

\bibitem{CMS-PAS-HIG-13-016}
CMS Collaboration,
\newblock CERN Report No. CMS-PAS-HIG-13-016, 2013 (unpublished).

\bibitem{Landau:1948kw}
L.~Landau,
\newblock Dokl.Akad.Nauk Ser.Fiz. {\bf 60}, 207 (1948).

\bibitem{Yang:1950rg}
C.-N. Yang,
\newblock Phys.Rev. {\bf 77}, 242 (1950).

\bibitem{ATLAS:2013vla}
ATLAS Collaboration,
\newblock CERN Report No. ATLAS-CONF-2013-031, 2013 (unpublished).

\bibitem{CMS-PAS-HIG-13-003}
CMS Collaboration,
\newblock CERN Report No. CMS-PAS-HIG-13-003, 2013 (unpublished).

\bibitem{Ellis:2012jv}
J.~Ellis, R.~Fok, D.~S. Hwang, V.~Sanz, and T.~You,
\newblock Eur.Phys.J. {\bf C73}, 2488 (2013), 1210.5229.

\bibitem{ATLAS-CONF-2013-013}
ATLAS Collaboration,
\newblock CERN Report No. ATLAS-CONF-2013-013, 2013 (unpublished).

\bibitem{Chatrchyan:2013mxa}
CMS Collaboration, S.~Chatrchyan {\em et~al.},
\newblock (2013), 1312.5353.

\bibitem{Choi:2002jk}
S.~Choi, D.~Miller, M.~Muhlleitner, and P.~Zerwas,
\newblock Phys.Lett. {\bf B553}, 61 (2003), hep-ph/0210077.

\bibitem{Buszello:2002uu}
C.~Buszello, I.~Fleck, P.~Marquard, and J.~van~der Bij,
\newblock Eur.Phys.J. {\bf C32}, 209 (2004), hep-ph/0212396.

\bibitem{Bredenstein:2006rh}
A.~Bredenstein, A.~Denner, S.~Dittmaier, and M.~Weber,
\newblock Phys.Rev. {\bf D74}, 013004 (2006), hep-ph/0604011.

\bibitem{Cao:2009ah}
Q.-H. Cao, C.~Jackson, W.-Y. Keung, I.~Low, and J.~Shu,
\newblock Phys.Rev. {\bf D81}, 015010 (2010), 0911.3398.

\bibitem{Gao:2010qx}
Y.~Gao {\em et~al.},
\newblock Phys.Rev. {\bf D81}, 075022 (2010), 1001.3396.

\bibitem{Englert:2010ud}
C.~Englert, C.~Hackstein, and M.~Spannowsky,
\newblock Phys.Rev. {\bf D82}, 114024 (2010), 1010.0676.

\bibitem{Ellis:2012wg}
J.~Ellis and D.~S. Hwang,
\newblock JHEP {\bf 1209}, 071 (2012), 1202.6660.

\bibitem{Ellis:2012mj}
J.~Ellis, V.~Sanz, and T.~You,
\newblock Phys.Lett. {\bf B726}, 244 (2013), 1211.3068.

\bibitem{Freitas:2012kw}
A.~Freitas and P.~Schwaller,
\newblock Phys.Rev. {\bf D87}, 055014 (2013), 1211.1980.

\bibitem{Coleppa:2012eh}
B.~Coleppa, K.~Kumar, and H.~E. Logan,
\newblock Phys.Rev. {\bf D86}, 075022 (2012), 1208.2692.

\bibitem{Plehn:2001nj}
T.~Plehn, D.~L. Rainwater, and D.~Zeppenfeld,
\newblock Phys.Rev.Lett. {\bf 88}, 051801 (2002), hep-ph/0105325.

\bibitem{Klamke:2007cu}
G.~Klamke and D.~Zeppenfeld,
\newblock JHEP {\bf 0704}, 052 (2007), hep-ph/0703202.

\bibitem{Godbole:2007cn}
R.~M. Godbole, D.~Miller, and M.~M. Muhlleitner,
\newblock JHEP {\bf 0712}, 031 (2007), 0708.0458.

\bibitem{Berge:2011ij}
S.~Berge, W.~Bernreuther, B.~Niepelt, and H.~Spiesberger,
\newblock Phys.Rev. {\bf D84}, 116003 (2011), 1108.0670.

\bibitem{Englert:2012ct}
C.~Englert, M.~Spannowsky, and M.~Takeuchi,
\newblock JHEP {\bf 1206}, 108 (2012), 1203.5788.

\bibitem{Djouadi:2013qya}
A.~Djouadi and G.~Moreau,
\newblock (2013), 1303.6591.

\bibitem{Harnik:2013aja}
R.~Harnik, A.~Martin, T.~Okui, R.~Primulando, and F.~Yu,
\newblock Phys.Rev. {\bf D88}, 076009 (2013), 1308.1094.

\bibitem{Ellis:2013yxa}
J.~Ellis, D.~S. Hwang, K.~Sakurai, and M.~Takeuchi,
\newblock JHEP {\bf 1404}, 004 (2014), 1312.5736.

\bibitem{ATLAS-CONF-2014-009}
The ATLAS collaboration,
\newblock CERN Report No. ATLAS-CONF-2014-009, 2014 (unpublished).

\bibitem{Chatrchyan:2014vua}
CMS Collaboration, S.~Chatrchyan {\em et~al.},
\newblock (2014), 1401.6527.

\bibitem{Low:2009di}
I.~Low, R.~Rattazzi, and A.~Vichi,
\newblock JHEP {\bf 1004}, 126 (2010), 0907.5413.

\bibitem{Espinosa:2012in}
J.~R. Espinosa, C.~Grojean, V.~Sanz, and M.~Trott,
\newblock JHEP {\bf 1212}, 077 (2012), 1207.7355.

\bibitem{Shifman:1979eb}
M.~A. Shifman, A.~Vainshtein, M.~Voloshin, and V.~I. Zakharov,
\newblock Sov.J.Nucl.Phys. {\bf 30}, 711 (1979).

\bibitem{Ellis:1975ap}
J.~R. Ellis, M.~K. Gaillard, and D.~V. Nanopoulos,
\newblock Nucl.Phys. {\bf B106}, 292 (1976).

\bibitem{Kniehl:1995tn}
B.~A. Kniehl and M.~Spira,
\newblock Z.Phys. {\bf C69}, 77 (1995), hep-ph/9505225.

\bibitem{Grazzini:2013mca}
M.~Grazzini and H.~Sargsyan,
\newblock JHEP {\bf 1309}, 129 (2013), 1306.4581.

\bibitem{Falkowski:2007hz}
A.~Falkowski,
\newblock Phys.Rev. {\bf D77}, 055018 (2008), 0711.0828.

\bibitem{Low:2010mr}
I.~Low and A.~Vichi,
\newblock Phys.Rev. {\bf D84}, 045019 (2011), 1010.2753.

\bibitem{Azatov:2011qy}
A.~Azatov and J.~Galloway,
\newblock Phys.Rev. {\bf D85}, 055013 (2012), 1110.5646.

\bibitem{Montull:2013mla}
M.~Montull, F.~Riva, E.~Salvioni, and R.~Torre,
\newblock Phys.Rev. {\bf D88}, 095006 (2013), 1308.0559.

\bibitem{Delaunay:2013iia}
C.~Delaunay, C.~Grojean, and G.~Perez,
\newblock JHEP {\bf 1309}, 090 (2013), 1303.5701.

\bibitem{Grojean:2013nya}
C.~Grojean, E.~Salvioni, M.~Schlaffer, and A.~Weiler,
\newblock JHEP {\bf 1405}, 022 (2014), 1312.3317.

\bibitem{Langenegger:2006wu}
U.~Langenegger, M.~Spira, A.~Starodumov, and P.~Trueb,
\newblock JHEP {\bf 0606}, 035 (2006), hep-ph/0604156.

\bibitem{Arnesen:2008fb}
C.~Arnesen, I.~Z. Rothstein, and J.~Zupan,
\newblock Phys.Rev.Lett. {\bf 103}, 151801 (2009), 0809.1429.

\bibitem{Bagnaschi:2011tu}
E.~Bagnaschi, G.~Degrassi, P.~Slavich, and A.~Vicini,
\newblock JHEP {\bf 1202}, 088 (2012), 1111.2854.

\bibitem{Azatov:2013xha}
A.~Azatov and A.~Paul,
\newblock JHEP {\bf 1401}, 014 (2014), 1309.5273.

\bibitem{Harlander:2013oja}
R.~V. Harlander and T.~Neumann,
\newblock Phys.Rev. {\bf D88}, 074015 (2013), 1308.2225.

\bibitem{Banfi:2013yoa}
A.~Banfi, A.~Martin, and V.~Sanz,
\newblock (2013), 1308.4771.

\bibitem{Englert:2013vua}
C.~Englert, M.~McCullough, and M.~Spannowsky,
\newblock Phys.Rev. {\bf D89}, 013013 (2014), 1310.4828.

\bibitem{Harlander:2013mla}
R.~V. Harlander, S.~Liebler, and T.~Zirke,
\newblock JHEP {\bf 1402}, 023 (2014), 1307.8122.

\bibitem{Plehn:2009rk}
T.~Plehn, G.~P. Salam, and M.~Spannowsky,
\newblock Phys.Rev.Lett. {\bf 104}, 111801 (2010), 0910.5472.

\bibitem{ATLAS2014ttH}
The ATLAS collaboration,
\newblock CERN Report No. ATLAS-CONF-2014-011, 2014 (unpublished).

\bibitem{CMS:2013fda}
CMS Collaboration,
\newblock CERN Report No. CMS-PAS-HIG-13-015, 2013 (unpublished).

\bibitem{CMS:2013sea}
CMS Collaboration,
\newblock CERN Report No. CMS-PAS-HIG-13-019, 2013 (unpublished).

\bibitem{CMS:2013tfa}
CMS Collaboration,
\newblock CERN Report No. CMS-PAS-HIG-13-020, 2013 (unpublished).

\bibitem{Artoisenet:2013vfa}
P.~Artoisenet, P.~de~Aquino, F.~Maltoni, and O.~Mattelaer,
\newblock Phys.Rev.Lett. {\bf 111}, 091802 (2013), 1304.6414.

\bibitem{Buckley:2013auc}
M.~R. Buckley, T.~Plehn, T.~Schell, and M.~Takeuchi,
\newblock JHEP {\bf 1402}, 130 (2014), 1310.6034.

\bibitem{Agashe:2004rs}
K.~Agashe, R.~Contino, and A.~Pomarol,
\newblock Nucl.Phys. {\bf B719}, 165 (2005), hep-ph/0412089.

\bibitem{Contino:2010rs}
R.~Contino,
\newblock (2010), 1005.4269.

\bibitem{Bellazzini:2014yua}
B.~Bellazzini, C.~Csáki, and J.~Serra,
\newblock (2014), 1401.2457.

\bibitem{Gillioz:2012se}
M.~Gillioz, R.~Grober, C.~Grojean, M.~Muhlleitner, and E.~Salvioni,
\newblock JHEP {\bf 1210}, 004 (2012), 1206.7120.

\bibitem{Chatrchyan:2013uxa}
CMS Collaboration, S.~Chatrchyan {\em et~al.},
\newblock Phys.Lett. {\bf B729}, 149 (2014), 1311.7667.

\bibitem{ATLAS:2013ima}
ATLAS Collaboration,
\newblock CERN Report No. ATLAS-CONF-2013-018, 2013 (unpublished).

\bibitem{TheATLAScollaboration:2013oha}
ATLAS collaboration,
\newblock CERN Report No. ATLAS-CONF-2013-056, 2013 (unpublished).

\bibitem{TheATLAScollaboration:2013jha}
ATLAS collaboration,
\newblock CERN Report No. ATLAS-CONF-2013-051, 2013 (unpublished).

\bibitem{deSimone:2012fs}
A.~De~Simone, O.~Matsedonskyi, R.~Rattazzi, and A.~Wulzer,
\newblock JHEP {\bf 1304}, 004 (2013), 1211.5663.

\bibitem{Aguilar-Saavedra:2013qpa}
J.~Aguilar-Saavedra, R.~Benbrik, S.~Heinemeyer, and M.~Perez-Victoria,
\newblock Phys.Rev. {\bf D88}, 094010 (2013), 1306.0572.

\bibitem{Buchkremer:2013bha}
M.~Buchkremer, G.~Cacciapaglia, A.~Deandrea, and L.~Panizzi,
\newblock Nucl.Phys. {\bf B876}, 376 (2013), 1305.4172.

\bibitem{Azatov:2013hya}
A.~Azatov, M.~Salvarezza, M.~Son, and M.~Spannowsky,
\newblock Phys.Rev. {\bf D89}, 075001 (2014), 1308.6601.

\bibitem{Kearney:2013oia}
J.~Kearney, A.~Pierce, and J.~Thaler,
\newblock JHEP {\bf 1308}, 130 (2013), 1304.4233.

\bibitem{Delaunay:2013pwa}
C.~Delaunay {\em et~al.},
\newblock JHEP {\bf 1402}, 055 (2014), 1311.2072.

\bibitem{Kunszt:1991qe}
Z.~Kunszt and F.~Zwirner,
\newblock Nucl.Phys. {\bf B385}, 3 (1992), hep-ph/9203223.

\bibitem{Barger:1991ed}
V.~D. Barger, M.~Berger, A.~Stange, and R.~Phillips,
\newblock Phys.Rev. {\bf D45}, 4128 (1992).

\bibitem{Baer:1991yc}
H.~Baer, M.~Bisset, C.~Kao, and X.~Tata,
\newblock Phys.Rev. {\bf D46}, 1067 (1992).

\bibitem{Gunion:1991er}
J.~F. Gunion and L.~H. Orr,
\newblock Phys.Rev. {\bf D46}, 2052 (1992).

\bibitem{Gunion:1991cw}
J.~F. Gunion, H.~E. Haber, and C.~Kao,
\newblock Phys.Rev. {\bf D46}, 2907 (1992).

\bibitem{Djouadi:1991tka}
A.~Djouadi, M.~Spira, and P.~Zerwas,
\newblock Phys.Lett. {\bf B264}, 440 (1991).

\bibitem{Spira:1995rr}
M.~Spira, A.~Djouadi, D.~Graudenz, and P.~Zerwas,
\newblock Nucl.Phys. {\bf B453}, 17 (1995), hep-ph/9504378.

\bibitem{Dawson:1990zj}
S.~Dawson,
\newblock Nucl.Phys. {\bf B359}, 283 (1991).

\bibitem{Graudenz:1992pv}
D.~Graudenz, M.~Spira, and P.~Zerwas,
\newblock Phys.Rev.Lett. {\bf 70}, 1372 (1993).

\bibitem{Haber:1984zu}
H.~E. Haber and G.~L. Kane,
\newblock Nucl.Phys. {\bf B250}, 716 (1985).

\bibitem{Aad:2013ija}
ATLAS, G.~Aad {\em et~al.},
\newblock JHEP {\bf 1310}, 189 (2013), 1308.2631.

\bibitem{TheATLAScollaboration:2013aia}
ATLAS collaboration,
\newblock CERN Report No. ATLAS-CONF-2013-068, 2013 (unpublished).

\bibitem{TheATLAScollaboration:2013xha}
ATLAS collaboration,
\newblock CERN Report No. ATLAS-CONF-2013-065, 2013 (unpublished).

\bibitem{TheATLAScollaboration:2013tha}
ATLAS collaboration,
\newblock CERN Report No. ATLAS-CONF-2013-061, 2013 (unpublished).

\bibitem{TheATLAScollaboration:2013gha}
ATLAS collaboration,
\newblock CERN Report No. ATLAS-CONF-2013-048, 2013 (unpublished).

\bibitem{ATLAS:2013pla}
ATLAS collaboration,
\newblock CERN Report No. ATLAS-CONF-2013-037, 2013 (unpublished).

\bibitem{ATLAS:2013bma}
ATLAS Collaboration,
\newblock CERN Report No. ATLAS-CONF-2013-025, 2013 (unpublished).

\bibitem{ATLAS:2013cma}
ATLAS Collaboration,
\newblock CERN Report No. ATLAS-CONF-2013-024, 2013 (unpublished).

\bibitem{Chatrchyan:2013xna}
CMS Collaboration, S.~Chatrchyan {\em et~al.},
\newblock Eur.Phys.J. {\bf C73}, 2677 (2013), 1308.1586.

\bibitem{CMS:2013ffa}
CMS Collaboration,
\newblock CERN Report No. CMS-PAS-SUS-13-014, 2013 (unpublished).

\bibitem{CMS:2013ida}
CMS Collaboration,
\newblock CERN Report No. CMS-PAS-SUS-13-008, 2013 (unpublished).

\bibitem{CMS:zrk}
CMS Collaboration,
\newblock CERN Report No. CMS-PAS-SUS-12-023, 2012 (unpublished).

\bibitem{Papucci:2014rja}
M.~Papucci, K.~Sakurai, A.~Weiler, and L.~Zeune,
\newblock (2014), 1402.0492.

\bibitem{Papucci:2011wy}
M.~Papucci, J.~T. Ruderman, and A.~Weiler,
\newblock JHEP {\bf 1209}, 035 (2012), 1110.6926.

\bibitem{Delgado:2012eu}
A.~Delgado, G.~F. Giudice, G.~Isidori, M.~Pierini, and A.~Strumia,
\newblock Eur.Phys.J. {\bf C73}, 2370 (2013), 1212.6847.

\bibitem{Kats:2011it}
Y.~Kats and D.~Shih,
\newblock JHEP {\bf 1108}, 049 (2011), 1106.0030.

\bibitem{Fan:2012jf}
J.~Fan, M.~Reece, and J.~T. Ruderman,
\newblock JHEP {\bf 1207}, 196 (2012), 1201.4875.

\bibitem{Fan:2011yu}
J.~Fan, M.~Reece, and J.~T. Ruderman,
\newblock JHEP {\bf 1111}, 012 (2011), 1105.5135.

\bibitem{Bai:2013xla}
Y.~Bai, A.~Katz, and B.~Tweedie,
\newblock JHEP {\bf 1401}, 040 (2014), 1309.6631.

\bibitem{Buchmuller:1985jz}
W.~Buchmuller and D.~Wyler,
\newblock Nucl.Phys. {\bf B268}, 621 (1986).

\bibitem{Grzadkowski:2010es}
B.~Grzadkowski, M.~Iskrzynski, M.~Misiak, and J.~Rosiek,
\newblock JHEP {\bf 1010}, 085 (2010), 1008.4884.

\bibitem{Giudice:2007fh}
G.~Giudice, C.~Grojean, A.~Pomarol, and R.~Rattazzi,
\newblock JHEP {\bf 0706}, 045 (2007), hep-ph/0703164.

\bibitem{Contino:2013kra}
R.~Contino, M.~Ghezzi, C.~Grojean, M.~Muhlleitner, and M.~Spira,
\newblock JHEP {\bf 1307}, 035 (2013), 1303.3876.

\bibitem{Elias-Miro:2013mua}
J.~Elias-Miro, J.~Espinosa, E.~Masso, and A.~Pomarol,
\newblock JHEP {\bf 1311}, 066 (2013), 1308.1879.

\bibitem{Pomarol:2013zra}
A.~Pomarol and F.~Riva,
\newblock JHEP {\bf 1401}, 151 (2014), 1308.2803.

\bibitem{Englert:2014uua}
C.~Englert {\em et~al.},
\newblock (2014), 1403.7191.

\bibitem{Bramante:2014gda}
J.~Bramante, A.~Delgado, and A.~Martin,
\newblock Phys.Rev. {\bf D89}, 093006 (2014), 1402.5985.

\bibitem{Ellis:2014dva}
J.~Ellis, V.~Sanz, and T.~You,
\newblock (2014), 1404.3667.

\bibitem{Dumont:2013wma}
B.~Dumont, S.~Fichet, and G.~von Gersdorff,
\newblock JHEP {\bf 1307}, 065 (2013), 1304.3369.

\bibitem{Mantler:2012bj}
H.~Mantler and M.~Wiesemann,
\newblock Eur.Phys.J. {\bf C73}, 2467 (2013), 1210.8263.

\bibitem{Baur:1989cm}
U.~Baur and E.~N. Glover,
\newblock Nucl.Phys. {\bf B339}, 38 (1990).

\bibitem{ATL-PHYS-PUB-2013-014}
CERN Report No. ATL-PHYS-PUB-2013-014, 2013 (unpublished).

\bibitem{Liss:1564937}
ATLAS Collaboration, A.~Liss and J.~Nielsen,
\newblock CERN Report No. ATL-PHYS-PUB-2013-007, 2013 (unpublished).

\bibitem{CMS-NOTE-2012-006}
CMS Collaboration,
\newblock CERN Report No. CMS-NOTE-2012-006. CERN-CMS-NOTE-2012-006, 2012
  (unpublished).

\bibitem{Alwall:2011uj}
J.~Alwall, M.~Herquet, F.~Maltoni, O.~Mattelaer, and T.~Stelzer,
\newblock JHEP {\bf 1106}, 128 (2011), 1106.0522.

\bibitem{Bahr:2008pv}
M.~Bahr {\em et~al.},
\newblock Eur.Phys.J. {\bf C58}, 639 (2008), 0803.0883.

\bibitem{Arnold:2012fq}
K.~Arnold {\em et~al.},
\newblock (2012), 1205.4902.

\bibitem{Bellm:2013lba}
J.~Bellm {\em et~al.},
\newblock (2013), 1310.6877.

\bibitem{Ellis:1987xu}
R.~K. Ellis, I.~Hinchliffe, M.~Soldate, and J.~van~der Bij,
\newblock Nucl.Phys. {\bf B297}, 221 (1988).

\bibitem{Harlander:2012hf}
R.~V. Harlander, T.~Neumann, K.~J. Ozeren, and M.~Wiesemann,
\newblock JHEP {\bf 1208}, 139 (2012), 1206.0157.

\bibitem{Boughezal:2013uia}
R.~Boughezal, F.~Caola, K.~Melnikov, F.~Petriello, and M.~Schulze,
\newblock JHEP {\bf 1306}, 072 (2013), 1302.6216.

\bibitem{Hoeche:2014lxa}
S.~Hoeche, F.~Krauss, and M.~Schonherr,
\newblock (2014), 1401.7971.

\bibitem{Dittmaier:2011ti}
LHC Higgs Cross Section Working Group, S.~Dittmaier {\em et~al.},
\newblock (2011), 1101.0593.

\bibitem{Anastasiou:2008tj}
C.~Anastasiou, R.~Boughezal, and F.~Petriello,
\newblock JHEP {\bf 0904}, 003 (2009), 0811.3458.

\bibitem{deFlorian:2009hc}
D.~de~Florian and M.~Grazzini,
\newblock Phys.Lett. {\bf B674}, 291 (2009), 0901.2427.

\bibitem{Bonvini:2014joa}
M.~Bonvini and S.~Marzani,
\newblock (2014), 1405.3654.

\bibitem{Mangano:2002ea}
M.~L. Mangano, M.~Moretti, F.~Piccinini, R.~Pittau, and A.~D. Polosa,
\newblock JHEP {\bf 0307}, 001 (2003), hep-ph/0206293.

\bibitem{Sjostrand:2006za}
T.~Sjostrand, S.~Mrenna, and P.~Z. Skands,
\newblock JHEP {\bf 0605}, 026 (2006), hep-ph/0603175.

\bibitem{Mangano:2001xp}
M.~L. Mangano, M.~Moretti, and R.~Pittau,
\newblock Nucl.Phys. {\bf B632}, 343 (2002), hep-ph/0108069.

\bibitem{Hoche:2006ph}
S.~Hoeche {\em et~al.},
\newblock (2006), hep-ph/0602031.

\bibitem{Nason:1987xz}
P.~Nason, S.~Dawson, and R.~K. Ellis,
\newblock Nucl.Phys. {\bf B303}, 607 (1988).

\bibitem{Beenakker:1988bq}
W.~Beenakker, H.~Kuijf, W.~van Neerven, and J.~Smith,
\newblock Phys.Rev. {\bf D40}, 54 (1989).

\bibitem{Moch:2008qy}
S.~Moch and P.~Uwer,
\newblock Phys.Rev. {\bf D78}, 034003 (2008), 0804.1476.

\bibitem{Cacciari:2011ma}
M.~Cacciari, G.~P. Salam, and G.~Soyez,
\newblock Eur.Phys.J. {\bf C72}, 1896 (2012), 1111.6097.

\bibitem{Dokshitzer:1997in}
Y.~L. Dokshitzer, G.~Leder, S.~Moretti, and B.~Webber,
\newblock JHEP {\bf 9708}, 001 (1997), hep-ph/9707323.

\bibitem{Wobisch:1998wt}
M.~Wobisch and T.~Wengler,
\newblock (1998), hep-ph/9907280.

\bibitem{Gripaios:2012th}
B.~Gripaios, K.~Nagao, M.~Nojiri, K.~Sakurai, and B.~Webber,
\newblock JHEP {\bf 1303}, 106 (2013), 1210.1938.

\bibitem{Katz:2010iq}
A.~Katz, M.~Son, and B.~Tweedie,
\newblock Phys.Rev. {\bf D83}, 114033 (2011), 1011.4523.

\bibitem{Dittmar:1997nea}
M.~Dittmar and H.~Dreiner,
\newblock  Report No. CMS-NOTE-1997-083, CERN-CMS-NOTE-1997-083, 1997
  (unpublished).

\bibitem{Dittmar:1996ss}
M.~Dittmar and H.~K. Dreiner,
\newblock Phys.Rev. {\bf D55}, 167 (1997), hep-ph/9608317.

\bibitem{Berger:2010xi}
C.~F. Berger, C.~Marcantonini, I.~W. Stewart, F.~J. Tackmann, and W.~J.
  Waalewijn,
\newblock JHEP {\bf 1104}, 092 (2011), 1012.4480.

\bibitem{Banfi:2012jm}
A.~Banfi, P.~F. Monni, G.~P. Salam, and G.~Zanderighi,
\newblock Phys.Rev.Lett. {\bf 109}, 202001 (2012), 1206.4998.

\bibitem{Buschmann:2014twa}
M.~Buschmann, C.~Englert, D.~Goncalves, T.~Plehn, and M.~Spannowsky,
\newblock Phys.Rev. {\bf D90}, 013010 (2014), 1405.7651.

\bibitem{Dolan:2014upa}
M.~J. Dolan, P.~Harris, M.~Jankowiak, and M.~Spannowsky,
\newblock (2014), 1406.3322.

\bibitem{Spannowsky:2013qb}
M.~Spannowsky and C.~Wymant,
\newblock Phys.Rev. {\bf D87}, 074004 (2013), 1301.0345.

\bibitem{Mellado200560}
B.~Mellado, W.~Quayle, and S.~L. Wu,
\newblock Physics Letters B {\bf 611}, 60  (2005).

\bibitem{Barr:2011he}
A.~J. Barr, S.~T. French, J.~A. Frost, and C.~G. Lester,
\newblock JHEP {\bf 1110}, 080 (2011), 1106.2322.

\bibitem{Gallicchio:2011xq}
J.~Gallicchio and M.~D. Schwartz,
\newblock Phys.Rev.Lett. {\bf 107}, 172001 (2011), 1106.3076.

\bibitem{ATLAS:2012zoa}
ATLAS Collaboration,
\newblock CERN Report No. ATLAS-CONF-2012-138, ATLAS-COM-CONF-2012-108, 2012
  (unpublished).

\bibitem{CMS:2013kfa}
CMS Collaboration,
\newblock CERN Report No. CMS-PAS-JME-13-002, 2013 (unpublished).

\bibitem{Larkoski:2013eya}
A.~J. Larkoski, G.~P. Salam, and J.~Thaler,
\newblock JHEP {\bf 1306}, 108 (2013), 1305.0007.

\bibitem{Mukhopadhyay:2014dsa}
S.~Mukhopadhyay, M.~M. Nojiri, and T.~T. Yanagida,
\newblock (2014), 1403.6028.

\bibitem{Lester:1999tx}
C.~Lester and D.~Summers,
\newblock Phys.Lett. {\bf B463}, 99 (1999), hep-ph/9906349.

\bibitem{Lester:2011nj}
C.~G. Lester,
\newblock JHEP {\bf 1105}, 076 (2011), 1103.5682.

\bibitem{Barr:2009mx}
A.~J. Barr, B.~Gripaios, and C.~G. Lester,
\newblock JHEP {\bf 0907}, 072 (2009), 0902.4864.

\bibitem{ATLAS-CONF-2013-030}
ATLAS Collaboration,
\newblock CERN Report No. ATLAS-CONF-2013-030, 2013 (unpublished).

\bibitem{Chatrchyan:2013iaa}
CMS Collaboration, S.~Chatrchyan {\em et~al.},
\newblock JHEP {\bf 1401}, 096 (2014), 1312.1129.

\bibitem{Junk:1999kv}
T.~Junk,
\newblock Nucl.Instrum.Meth. {\bf A434}, 435 (1999), hep-ex/9902006.

\end{thebibliography}

\end{document}